\documentclass[11pt]{article}
\usepackage{mymacros}

\title{Evolution With(out) Time: Relational Holography \& BPS Complexity Growth in $\mathcal{N}=2$ Double-Scaled SYK}
\author[a]{Sergio E. Aguilar-Gutierrez\orcidlink{0000-0003-0308-0061}}
\affiliation[a]{Qubits and Spacetime Unit, Okinawa Institute of Science and Technology Graduate University,\footnote{\begin{CJK}{UTF8}{min}沖縄科学技術大学院大学\end{CJK}} 1919-1 Tancha, Onna, Okinawa 904 0495, Japan}
\emailAdd{sergio.ernesto.aguilar@gmail.com}

\abstract{How do we describe non-trivial bulk measurements relative to an observer (i.e.~relationally) when both the observer and the system it probes may/may not evolve in time? How can we interpret this holographically; particularly for zero-energy BPS states in supersymmetric theories? We address these questions, in the $\mathcal{N}=2$ double-scaled SYK model and its putative bulk dual by: (i) formulating a holographic procedure in the language of quantum reference frames to gravitationally dress bulk observables to ``clocks'' parametrized by both boundary time and R-charge; and (ii) proposing a \emph{new measure of Krylov complexity} with R-charge in the boundary theory that probes zero-energy BPS states. Holographically, this proposal reproduces a relational bulk observable, a BPS wormhole length. We contrast this to the Krylov complexity for Hartle-Hawking states with non-trivial time flow. The latter reproduces the same observable as for the bosonic DSSYK in the semiclassical limit, while its quantum fluctuations can capture supersymmetric corrections {{depending on the specific initial state}}.}

\begin{document}

\maketitle
\section{Introduction}\label{sec:intro}
\paragraph{Relational Holography}Generally, the Hilbert space of a holographic boundary theory is isomorphic to the physical Hilbert space (where diffeomorphism and other gauge constraints are imposed) of its dual gauge-invariant\footnote{By gauge-invariant we mean both diffeomorphism and internal gauge symmetry invariance.} gravity theory. Thus, while observables in the bulk theory are relational (i.e.~the operators belonging to the gauge-invariant algebra are defined relative to internal degrees of freedom within the system)\footnote{In holographic settings, observables are usually dressed with respect to the asymptotic or finite boundary cutoff of the spacetime when there is one (see e.g.~\cite{Mertens:2025rpa,Blommaert:2019hjr,Mertens:2019bvy,Blommaert:2020yeo,Blommaert:2020seb,Harlow:2021dfp,Nitti:2024iyj} in JT gravity), which is an example of an inmaterial frame, in the terminology of \cite{Carrozza:2022xut}. Alternatively, one might also implement the dressing with respect to matter within the bulk, a material frame \cite{Carrozza:2022xut}, e.g.~an infalling observer into a black hole \cite{Jafferis:2020ora,Gao:2021tzr,deBoer:2022zps,Aguilar-Gutierrez:2023odp,DeVuyst:2022bua}, or Goldstone boson modes in bath-AdS systems \cite{Geng:2025bcb}.}; the corresponding boundary theory observables are not relational {{with respect to gauge frames}}, in the sense that the operators acting on the boundary Hilbert space belong to a gauge-invariant algebra\footnote{One may additionally impose physical constraints instead of gauge ones to construct the physical Hilbert space. I thank Josh Kirklin for comments.} without dressing them to a subsystem in the boundary theory. Thus, we define \emph{relational holography}\footnote{This differs from our previous work \cite{Aguilar-Gutierrez:2025hty} based on \cite{Narovlansky:2023lfz} which has two-boundary theories instead.} as a framework to describe the entries in the holographic dictionary with relational observables in the bulk, which are not relational in the boundary side.

The definitions above rely on the existence of a reference frame with respect to which time measurements are made in order to obtain (dressed) observables, such as in the Page-Wootters (PW) mechanism \cite{Page:1983uc,Wootters:1984wfv}. {{In this formalism,}} the physical Hilbert space {{with respect to a given QRF is obtained from}} an isomorphism to the {{gauge-invariant}} Hilbert space {{(i.e.~after implementing gauge constraints) by applying a reduction map that involves the projecting onto states characterizing the QRF. In this way,}} the PW reduction can be implemented in the perspective neutral approach to quantum reference frames (QRFs) \cite{Krumm:2020fws,Hohn:2017cpr,Hoehn:2019fsy,Hoehn:2020epv,Hoehn:2023ehz,Vanrietvelde:2018pgb,Vanrietvelde:2018dit,delaHamette:2021oex,Hohn:2018toe,Hohn:2018iwn,Hoehn:2021flk,Giacomini:2021gei,Yang_2020,DeVuyst:2024uvd,DeVuyst:2024pop,AliAhmad:2024qrf,AliAhmad:2024vdw,AliAhmad:2024wja,Fewster:2024pur,DeVuyst:2025ezt,Araujo-Regado:2025ejs,Araujo-Regado:2024dpr} to adopt a fixed observer perspective {{describing the remainder of the system with respect to the QRF}}. {In most examples in the literature the state of the observer is specified by their proper time \cite{Krumm:2020fws,Hohn:2017cpr,Hoehn:2019fsy,Hoehn:2020epv,Hoehn:2023ehz,Vanrietvelde:2018pgb,Vanrietvelde:2018dit,delaHamette:2021oex,Hohn:2018toe,Hohn:2018iwn,Hoehn:2021flk,Giacomini:2021gei,DeVuyst:2024uvd,DeVuyst:2024pop,DeVuyst:2025ezt,Araujo-Regado:2025ejs}.} However, there are physical theories that may or may not experience time flow,\footnote{This means that there are no non-trivial physical frame reorientations in the QRF sense (see e.g.~\cite{DeVuyst:2024uvd} for a didactical explanation) within the same U$(1)_R$ symmetry sector.} such as the zero-energy ground states in supersymmetric (SUSY) theories (or any energy eigenstate more generally). {{Regardless}}, there is still a non-trivial R-charge functional dependence for states and operators that is associated to spectral chaos in the fortuity program (see e.g.~\cite{Chang:2024lxt,Chang:2024zqi,Chang:2025rqy,Chen:2024oqv,deMelloKoch:2025cec,Hughes:2025car}). More general states will have both boundary time (parametrizing the gauge orbits of the system being probed) and R-charge (parametrizing a physical U$(1)_R$ symmetry) dependence.\footnote{{The R-charge is localized in the (non-gravitating) boundary of the bulk spacetime.}} It is therefore interesting to formulate gravitational dressings when the boundary time may or may not trivialize, as the procedure is only known when there is non-trivial observer time (taken as the boundary time in this setting). Moreover, so far the QRF approach to SUSY theories has so far been overlooked to the best of the author's knowledge.\footnote{{{There is a related technique, the dressing field method (see \cite{Francois:2024rdm} for a review) for defining different types of relational dressings, that has been applied to supersymmetric theories, see e.g.~\cite{Francois:2024laf,Francois:2024xqi,Francois:2025nnk,Francois:2025odk}, albeit not in the QRF formalism.}}}

In this work, we develop bulk relational observables and define Krylov complexity for boundary theory states (spread complexity \cite{Balasubramanian:2022tpr}) in a SUSY holographic setting with or without QRF time evolution. We specialize most arguments to the $\mathcal{N}=2$ double-scaled Sachdev–Ye–Kitaev (DSSYK) model \cite{Berkooz:2020xne,Boruch:2023bte}\footnote{See \cite{Sachdev_1993,kitaevTalks1,kitaevTalks2,kitaevTalks3} for original work on the SYK model; \cite{Fu:2016vas} for its $\mathcal{N}=2$ generalization; \cite{Berkooz:2018jqr,Berkooz:2018qkz,Erd_s_2014,Cotler:2016fpe} for original work on the bosonic DSSYK model; \cite{Berkooz:2024lgq} for a recent review; and \cite{Blommaert:2023opb,Belaey:2025ijg} for other developments in the SUSY case.} for concreteness and to illustrate the main concepts explicitly in a solvable setting. We are particularly interested in studying a symmetry sector within the (super-chord \cite{Boruch:2023bte}) Hilbert space that is annihilated by all the supercharges. This results in an exactly zero-energy state {{in this model}}, denoted Bogomol’nyi– Prasad– Sommerfield (BPS) \cite{Witten:1978mh} (see a review in \cite{moore2010pitp}) Hartle-Hawking \cite{Hartle:1983ai} (HH)  state in \cite{Boruch:2023bte}.\footnote{In general, BPS states preserve a fraction of the SUSY, and they may not have trivial evolution. We comment how that is realized in the set-up in this work in Sec.~\ref{ssec:1/2 BPS wormholes}.} While any energy eigenstate in a general physical system satisfying the Schrödinger equation does not evolve; the HH BPS state has a non-trivial R-charge dependence and it is expected to be dual to the BPS wormhole \cite{Boruch:2023bte}. One of the main advantages of the relational framework in this setting is that based on a few postulates\footnote{These assumptions are input from holography including a correspondence between the Wheeler-DeWitt (WDW) and Schrödinger equations in the bulk and boundary theories respectively, and the isomorphism between the bulk physical Hilbert space with the boundary Hilbert space, as discussed in Sec.~\ref{sec:Relational Holo}.} one can recover a gravitationally dressed operators and perspective reduced physical states in the bulk theory with respect to an observer {{(which corresponds to the asymptotic boundary in this setting)}} from information about the boundary theory, even while the latter is not relational in the sense introduced above.

\paragraph{Krylov (Spread) Complexity \& DSSYK Model}
We now discuss specific boundary observables. Krylov operator \cite{Parker:2018yvk} and spread complexity \cite{Balasubramanian:2022tpr} (see \cite{Baiguera:2025dkc,Nandy:2024htc,Rabinovici:2025otw} for recent reviews{{, and Sec.~\ref{sec:non BPS N2} for details about the definition}}) are measures expected to discriminate integrable and chaotic systems \cite{Bhattacharjee:2024yxj,Nandy:2024mml,Bhattacharya:2023zqt,Aguilar-Gutierrez:2025hbf,Huh:2024ytz,Baggioli:2025knt,Gill:2025upp,Fu:2024fdm,Baggioli:2024wbz,Alishahiha:2024vbf,Rabinovici:2022beu};\footnote{However, Krylov operator complexity is not always a reliable chaos measure \cite{Chapman:2024pdw,Avdoshkin:2022xuw,Camargo:2022rnt} {{(see also \cite{Bhattacharjee:2022vlt} for early, and \cite{Aguilar-Gutierrez:2025hbf} late time analysis of the reliability of Krylov operator complexity in an integrable system example)}}.} Krylov complexity in the DSSYK model is intricately connected to scrambling dynamics \cite{Ambrosini:2024sre}; chaos as measured by out-of-time-ordered correlators (OTOCs) \cite{Aguilar-Gutierrez:2025hty,Aguilar-Gutierrez:2025mxf,Aguilar-Gutierrez:2025pqp}; and it is a well-established entry in its holographic dictionary, where it generically manifests as a wormhole length \cite{Rabinovici:2023yex,Heller:2024ldz,Lin:2023trc,Aguilar-Gutierrez:2025pqp,Aguilar-Gutierrez:2025mxf,Aguilar-Gutierrez:2024nau,Aguilar-Gutierrez:2025hty,Xu:2024hoc,Balasubramanian:2024lqk} in the bulk.\footnote{There are different proposals for the bulk dual of the DSSYK model (which may be compatible with each other \cite{Aguilar-Gutierrez:2025hty,Blommaert:2025eps,Aguilar-Gutierrez:2024oea}) beyond the low energy limit, including sine dilaton gravity \cite{Blommaert:2023opb,Blommaert:2024whf,Blommaert:2024ymv,Blommaert:2025avl,Blommaert:2025rgw,Bossi:2024ffa,Blommaert:2025eps,Cui:2025sgy,Aguilar-Gutierrez:2025pqp,Aguilar-Gutierrez:2024oea} (which is related to complex Liouville string \cite{Collier:2024kmo,Collier:2024kwt,Collier:2024lys,Collier:2024mlg,Collier:2025lux,Collier:2025pbm}) and three-dimensional de Sitter (dS$_3$) space in stretched horizon holography \cite{Susskind:2021esx,Susskind:2022bia,Susskind:2023hnj,Lin:2022nss,Rahman:2022jsf,Rahman:2023pgt,Rahman:2024iiu,Rahman:2024vyg,Sekino:2025bsc,Miyashita:2025rpt}, and static patch solipsism \cite{Anninos:2011af} approaches \cite{Narovlansky:2023lfz,Verlinde:2024znh,Verlinde:2024zrh,Narovlansky:2025tpb,Aguilar-Gutierrez:2024nau,Blommaert:2025eps}. Other relations between dS$_3$ space and a single DSSYK model can be found in \cite{Milekhin:2023bjv,Okuyama:2025hsd,Yuan:2024utc,Gaiotto:2024kze,Tietto:2025oxn}. Thus, these developments generically point towards a ultraviolet (UV) finite quantum cosmology model as the bulk dual theory.} Other approaches regarding OTOCs in the bosonic DSSYK can be found in \cite{Berkooz:2018jqr,Lin:2023trc,Narovlansky:2025tpb,Berkooz:2022fso}; and Krylov complexity in the DSSYK and related models in \cite{Bhattacharjee:2022ave,Anegawa:2024yia,Miyaji:2025ucp,Nandy:2024zcd}. 

In contrast to the bosonic case, the $\mathcal{N}=2$ DSSYK model \cite{Berkooz:2020xne,Boruch:2023bte} remains vastly underexplored in the literature, particularly regarding chaotic measures when there is trivial time evolution, and the emergence of the bulk dual theory. A reason to be interested in this setting is that the $\mathcal{N}=2$ DSSYK is a UV finite completion of the corresponding Jackiw-Teitelboim (JT) \cite{JACKIW1985343,TEITELBOIM198341} supergravity (e.g.~\cite{Penington:2024sum,Lin:2022zxd,Lin:2022rzw,Turiaci:2023jfa}) from a boundary perspective \cite{Boruch:2023bte}. The super-Schwarzian describes fluctuations in the near horizon region of near extremal supergravity black holes in higher dimensions \cite{Boruch:2023trc,Boruch:2022tno,Heydeman:2020hhw,Lin:2025wof}, and it dominates over other string theoretic corrections \cite{Chang:2024lxt} at low energies. Even though the boundary model is an ensemble-averaged description of a theory with infinite particles, one would expect that it retains relevant features to describe more general UV complete models of quantum gravity with BPS states, at least up to some regime. 

To deduce the relationship between wormhole geodesics in the bulk and chord number in the boundary, one might consider the SUSY enhancement in the $\mathcal{N}=2$ DSSYK to a $\mathcal{N}=4$ superalgebra once matter insertions are added \cite{Boruch:2023bte}. One then needs to distinguish between left and right sides of the corresponding chord diagram, where each side is associated with a $\mathcal{N}=2$ theory. However, by considering the limit of vanishing conformal dimension for the matter insertion, the analysis of the boundary BPS state dual to a BPS wormhole in SUSY two-dimensional anti-de Sitter (AdS$_2$) black hole is simplified. The length of the BPS wormhole in the bulk can then be matched to the expectation value of the total chord number in the BPS HH state  \cite{Boruch:2023bte}. 

\paragraph{Purpose of This Work}The $\mathcal{N}=2$ DSSYK is a natural laboratory to explore probes of BPS and non-BPS states, as well as to study their relational interpretation in the putative bulk dual theory. For instance, based on the bosonic case, one would expect that the length of a BPS wormhole (which is gravitationally dressed), and the chord number in the boundary theory are related to (some notion) of Krylov complexity \cite{Rabinovici:2023yex,Heller:2024ldz,Aguilar-Gutierrez:2025pqp,Xu:2024gfm,Aguilar-Gutierrez:2025hty,Aguilar-Gutierrez:2025mxf,Balasubramanian:2024lqk}. However, the original definition relies on Liovillian \cite{Parker:2018yvk} or Hamiltonian evolution \cite{Balasubramanian:2022tpr}\footnote{There are generalizations where one can include other generators that do not need to be related to time evolution, which we comment about in Sec.~\ref{ssec:BPS Krylov proposal}.}. Krylov complexity is then trivial when there is no time flow, which is our case of interest. Nevertheless, zero-energy BPS states are still expected to be chaotic according to spectral measures of chaos  \cite{Chang:2024lxt,Chang:2024zqi,Chang:2025rqy,Chen:2024oqv,deMelloKoch:2025cec,Hughes:2025car} that distinguish typical black hole microstates from horizonless geometries using supercharge cohomology \cite{Chang:2024zqi}. In contrast to the problem of time in gravity (see e.g.~\cite{Isham:1992ms}), where one can incorporate an internal degree of freedom (i.e.~an arbitrary subsystem) to define the evolution of the rest of the system relative to it, there is, seemingly, no auxiliary observer measuring non-trivial time flow that can be incorporated in zero-energy BPS systems (although we comment on some alternatives in Sec.~\ref{ssec:outlook}). Given that both the expectation value of the total chord number and wormhole length have non-trivial R-charge dependence, it is natural to expect that the ``clock'' observer (a QRF) should also measure R-charge to appropriately describe the BPS and non-BPS systems. These observations motivate us to develop relational holography and spread complexity for BPS and more general states for the $\mathcal{N}=2$ DSSYK model. 

In the first part of this study, we address a general problem
\begin{quote}
    \emph{How do we describe the different sectors within the physical Hilbert space that may or may not lack a time flow to define non-trivial dressed bulk observables? and what do they correspond to from the boundary side?}
\end{quote}
As we argue, there is a natural extension of the PW mechanism that allows to describe all physical states, including exactly zero-energy BPS states, and recover non-trivial observables. The main input is the R-charge dependence in the states and/or operator, as well as time dependence if there is one. While the details of the construction are based on the bulk interpretation of the (super-)chord Hilbert space in \cite{Lin:2022rbf,Boruch:2023bte} for the $\mathcal{N}=2$ DSSYK model, and physical symmetries in sine dilaton gravity \cite{Blommaert:2024whf}; we expect that the general arguments can be used in more generic SUSY theories with R-charge.

In the second part of this work, which can be read independently of the previous one but it is motivated by it, we study and develop extensions of spread complexity in the boundary theory, and its bulk interpretation as a relational observable. Our guiding question is:
\begin{quote}
    \emph{Is there a boundary measure of complexity of the $\mathcal{N}=2$ DSSYK model that captures the dual BPS wormhole length evolution in the BPS sector?}
\end{quote}
{{We answer this question affirmatively.}} The only other type of ``evolution'', in the sense of functional dependence with respect to some parameter, for zero-energy BPS states is in terms of the R-charge.\footnote{In contrast, in a different approach from ours, symmetry resolved Krylov and spread complexity \cite{Caputa:2025mii,Caputa:2025ozd}, one would separate the different R-charges as symmetry sectors to evaluate a time dependent Krylov complexity, while here the R-charge takes the role of time itself. It would be interesting to investigate possible connections with the other approaches further; see comments about this in Sec.~\ref{ssec:BPS Krylov proposal}.} We introduce a natural extension of spread complexity that measures wavefunction spreading for BPS states in terms of the R-charge instead of physical time.\footnote{This should be differentiated from other approaches, like generalized Krylov complexity \cite{FarajiAstaneh:2025thi} where one might instead use the R-charge generator in place of the Hamiltonian. In contrast, our approach works directly from the constraints imposed by the $\mathcal{N}=4$ charges in the HH BPS state. It might be interesting to make a direct comparison with the previous approach in this setting.} This extension of spread complexity reproduces the expectation value of the chord number in the BPS state, which is known to match in the semiclassical limit to a wormhole length in $\mathcal{N}=2$ JT supergravity \cite{Boruch:2023bte}. Thus, this proposal is an entry in the holographic dictionary of the $\mathcal{N}=2$ DSSYK. Given that the BPS wormhole length is one of the canonical variables of the super-Schwarzian formulation of JT-supergravity \cite{Lin:2022rzw,Lin:2022zxd}, it might play a role in formulating the corresponding dual supergravity theory Hamiltonian of the $\mathcal{N}=2$ DSSYK beyond low energies, and in understanding chaos, or the lack of,\footnote{For instance, the bosonic DSSYK is submaximally chaotic with respect to the chaos bound \cite{Maldacena:2015waa} as measured by OTOCs \cite{Berkooz:2018jqr,Aguilar-Gutierrez:2025mxf,Aguilar-Gutierrez:2025pqp,Lin:2023trc}. As seen in App.~\ref{app:thermo}, the semiclassical thermodynamics is similar to the bosonic case, so one might expect it is also submaximally chaotic.} in the boundary theory.

Besides the BPS state above, interpreted as a HH preparation of a two-sided AdS$_2$ BPS-black hole in the bulk \cite{Boruch:2023bte}, there are other notions of HH states with non-trivial time dependence. For instance, there are orthogonal bosonic subspaces, where we construct HH states respect to each of them; and a HH state prepared by complex time evolving the maximally entangled state for a fixed R-charge, which encodes information about all the spectrum (including BPS states) \cite{Berkooz:2020xne}. This state is used to define the thermal ensembles in the theory (see App.~\ref{app:thermo}). However, in both of the cases above, the standard definition of spread complexity (see Sec.~\ref{sec:non BPS N2} for details) in the semiclassical limit leads to similar results as the bosonic DSSYK \cite{Rabinovici:2023yex}, as expected from a bulk analysis in \cite{Mertens:2022irh}. Yet, there are quantum corrections in the Krylov basis and spread complexity that contain information about the deviations from the purely bosonic case (see Sec.~\ref{ssec:TFD wormhole}).

We provide a brief overview of the results on complexity growth in Tab.~\ref{tab:structure}.
\begin{table}[t!]
    \centering
    \begin{tabular}{cc}\Xhline{2\arrayrulewidth}
    \textbf{Reference state}&\textbf{Wormhole length}\\
    \textbf{in spread complexity}&\\\Xhline{2\arrayrulewidth}
    BPS HH state & BPS wormhole \\
    &Sec.~\ref{sec:BPS wormholes}\\\hline
Bosonic subpaces & Bosonic wormhole\\
HH states&Sec.~\ref{ssec:bosonic subspaces}\\\hline
Maximally entangled& Bosonic wormhole\\
HH state&Sec.~\ref{ssec:TFD wormhole}\\
     \end{tabular}
    \caption{Different reference states in the evaluation of spread complexity in the $\mathcal{N}=2$ DSSYK model; and the corresponding observable in $\mathcal{N}=2$ JT supergravity in the semiclassical limit. In all cases the dual observables are wormhole lengths, which in some cases lead to the same answer as the bosonic theory in \cite{Rabinovici:2023yex}. Similar results are recovered for the $\mathcal{N}=1$ case in App.~\ref{app:N1 wormholes}.}\label{tab:structure}
    \end{table}

\paragraph{Plan of the Paper}
In Sec.~\ref{sec:set up N2} we briefly review background material on the $\mathcal{N}=2$ DSSYK model. In Sec.~\ref{sec:Relational Holo} we study the bulk interpretation of the super-chord Hilbert space in terms of gravitational dressings to define diffeomorphism-invariant observables with or without boundary time evolution, including the relevant one for spread complexity in the boundary. In Sec.~\ref{sec:BPS wormholes} we propose a definition for the spread complexity of BPS states in the model. It reproduces the expectation value of the total chord number in the same state without approximations, and, in the semiclassical limit, it matches with a wormhole length in $\mathcal{N}=2$ JT supergravity in the corresponding state. In Sec.~\ref{sec:non BPS N2} we study the spread complexity of non-BPS states (which has a non-trivial time dependence). We find agreement with wormhole lengths in JT supergravity. We conclude with a discussion and future directions in Sec.~\ref{sec:discussion}. 

We also include several appendices with technical details and other aid for the reader. In App.~\ref{app:notation} we summarize the notation (acronyms and the different symbols) used throughout the paper. In App.~\ref{app:extra background} we provide complementary background to Sec.~\ref{sec:set up N2}. In App.~\ref{app:N1 wormholes}, we do similar calculations as in the main text for $\mathcal{N}=1$ (instead of $\mathcal{N}=2$) DSSYK, i.e.~defining an extension for the spread complexity of BPS states, and we study the standard definition of spread complexity for a non-BPS HH state. Also, since most of the notation throughout this work follows that in \cite{Boruch:2023bte}, while the normalizations are chosen as in \cite{Berkooz:2020xne} for convenience; in
App.~\ref{app:BLY BBNR} we include lighting comparison between the normalizations in \cite{Berkooz:2020xne} and \cite{Boruch:2023bte}. In App.~\ref{app:thermo} we study the semiclassical limit of the 
$\mathcal{N}=2$ DSSYK partition function, and its triple-scaling limit. In App.~\ref{app:alternative} we work on an alternative definition of spread complexity constructed from the Krylov basis of the effective Hamiltonian \eqref{eq:BPS HH state j}. Later, App.~\ref{app:detail ortho eval} contains technical steps on the evaluation of spread complexity in Sec.~\ref{ssec:bosonic subspaces}. Similarly, App.~\ref{app:spread zero chord} contains details relevant for Sec.~\ref{ssec:TFD wormhole}. Meanwhile, in App.~\ref{app:alternative basis} we provide more details about the basis for the Hamiltonian explored in Sec.~\ref{ssec:TFD wormhole}.

\section{Brief Review of \texorpdfstring{$\mathcal{N}=2$}{} Double-Scaled SYK}\label{sec:set up N2}
In this section we provide some background material on the $\mathcal{N}=2$ DSSYK, including the Hilbert space construction in Sec.~\ref{ssec:aux system N2}, and BPS states in Sec.~\ref{ssec: BPS wormhole}.

\subsection{Super-Chord Hilbert space}\label{ssec:aux system N2}

The $\mathcal{N}=2$ SYK was introduced in \cite{Fu:2016vas}; its double-scaled limit was constructed in \cite{Berkooz:2020xne}, and it was further developed in \cite{Boruch:2023bte} (see other developments in e.g.~\cite{Blommaert:2023opb,Belaey:2025ijg}).

The super-chord Hilbert space of the $\mathcal{N}=2$ DSSYK is constructed from the following states (see App.~\ref{app:extra background} for more details)
\begin{equation}\label{eq:physical state N2 jR}
    \mH_{\rm super-chord}=\qty{\ket{\Omega,j},~\ket{n,XO,j},~\ket{n,OX,j},~\ket{n,XX,j},~\ket{n,OO,j}}_{n\in\mathbb{N},~j\in\mathbb{Z}}~,
\end{equation}
where the labels $X$ and $O$ represent two types of nodes in an oriented chord diagram, where $n$ indicates the number of pairs of $X$ and $O$ nodes, which gives rise to: bosonic states built from operators products of the type $XO\dots XO$ (labeled $XO$) and $OX\dots OX$ ($OX$); while $XO\dots XOX$ ($XX$) and $OX\dots OXO$ ($OO$) for the fermionic states; and $j$ is the R-charge. See below \eqref{eq:H aux} for details. Meanwhile, $\ket{\Omega,j}$ represents the maximally entangled state of the model for a fixed R-charge sector.

\paragraph{Maximally Entangled State \& Bosonic Subspaces}
There are bosonic states generated by the Hamiltonian acting on $\ket{\Omega,j}$,\footnote{On the other hand, the fermionic states $\qty{\ket{n,OO,j},~\ket{n,XX,j}}$ are not affected by the DSSYK Hamiltonian \cite{Berkooz:2020xne}, so they will not play a major role in the discussion of the Krylov space (although one indeed has to incorporate them to deduce the BPS HH state \cite{Boruch:2023bte}).} which take the form \cite{Berkooz:2020xne},\footnote{This differs from \cite{Berkooz:2020xne} by an overall scaling in the Hamiltonian by a factor $\sqrt{q}$.} 
\begin{subequations}\label{eq:H N2 antisymmetric form}
\begin{align}
    &\hH\ket{\Omega,j}=q^{-1/2}k\qty(q^{-j_R}\ket{H_0}+q^{j_R}\ket{\bar H_0})~,\label{eq:H N2 antisymmetric form_a}\\
    &\hH\ket{H_n}=q^{-1/2}k\qty(\ket{H_{n+1}}+\qty(1-q^{2n})\ket{H_{n-1}}+\qty(q^{-j_R+1/2}+q^{j_R-1/2})\ket{H_n})~,\\
    &\hH\ket{\bar H_n}=q^{-1/2}k\qty(\ket{\bar H_{n+1}}+\qty(1-q^{2n})\ket{\bar H_{n-1}}+\qty(q^{j_R+1/2}+q^{-j_R-1/2})\ket{\bar H_n})~,
\end{align}
\end{subequations}
where $q:=\rme^{-\lambda}\in[0,1)$ with $\lambda$ a fixed parameter (see \eqref{eq:double scale}), while
\begin{equation}\label{eq;def jR}
    j_R:=-j/2,\quad j\in\mathbb{Z}~,
\end{equation}
while $k$ is an overall constant, and the basis is given by
\begin{subequations}\label{eq:ortho bosonic basis}
\begin{align}
    \ket{H_0}&:=q^{j_R}\ket{XO,j}+\ket{\Omega,j}~,\\
    \ket{H_{n\geq1}}&:=\hmQ_R\ket{n,XX,j+1}=q^n\ket{n,OX,j}+\ket{n,XO,j}+q^{j_R}\ket{n+1,XO,j}~,\\
    \ket{\bar H_0}&:=q^{-j_R}\ket{OX,j_R}+\ket{\Omega,j}~,\\
    \ket{\bar H_{n\geq1}}&:=\hmQ_R^\dagger\ket{n,OO,j-1}=q^n\ket{n,XO,j}+\ket{n,OX,j}+q^{-j_R}\ket{n+1,OX,j}~.
\end{align}
\end{subequations}
Note $\ket{H_0}$ and $\ket{\bar H_0}$ are different states with respect to the zero-chord state $\ket{\Omega,j}$.

\subsection{BPS Wormhole Length from Chord Number}\label{ssec: BPS wormhole}
One can construct the HH BPS state in $\mathcal{N}=2$ DSSYK \cite{Boruch:2023bte}, which we label $\ket{\Psi,j}$ by demanding it annihilates all the supercharges
\begin{equation}\label{eq:supercharge constraints}
    \hmQ_{L/R}\ket{\Psi,j}=\hmQ^\dagger_{L/R}\ket{\Psi,j}=0~,
\end{equation}
which describes the ground state of the model. The solution of the supercharge constraints above can be expressed as\footnote{We stress we are using normalization of states in \cite{Berkooz:2020xne} instead of \cite{Boruch:2023bte} in the expression below. Note also there is no fermionic superpartner for this state in this one-dimensional model, while there can be one in more general theories.}
\begin{equation}\label{eq:BPS HH state j}
    \ket{\Psi,j}=\sum_{n=0}^\infty (\alpha_n\ket{n,XO,j}+\beta_n\ket{n,OX,j})~,
\end{equation}
where the coefficients obey a recurrence relation\footnote{The solution with $\alpha_0=\beta_0=1$ is \cite{Boruch:2023bte}\begin{equation}\label{eq:alpha bbeta n}
\begin{aligned}
    &\alpha_n=\frac{q^{3n/2}}{(q^2;q^2)_n}H_n\qty(-\eval{\cosh(\lambda\qty(j_R+\frac{1}{2}))}q^2)~,\\
    &\beta_n=\frac{q^{3n/2}}{(q^2;q^2)_n}H_n\qty(-\eval{\cosh(\lambda\qty(j_R-\frac{1}{2}))}q^2)~,
\end{aligned}
\end{equation}
where $(a;q)_n=\prod_{k=0}^{n-1}(1-aq^k)$ is the q-Pochhammer symbol, $H_n(x|q)$ the q-Hermite polynomials \eqref{eq:H_n def}.}
\begin{subequations}\label{eq:coefficients alpha beta}
    \begin{align}
    (q^3-q^{2n-1})\alpha_{n+1}+\qty(q^{-j_R-2}+q^{j_R-1})\alpha_n+\alpha_{n-1}=&0~,\\
    (q^3-q^{2n-1})\beta_{n+1}+\qty(q^{j_R-2}+q^{-j_R-1})\beta_n+\beta_{n-1}=&0~.
\end{align}
\end{subequations}
One may compute the two-point correlation function of uncharged matter chords (\ref{eq:matter chord op}), which acts as the generator of the total chord number \cite{Boruch:2023bte}:
\begin{equation}\label{eq:ell j}
    \begin{aligned}
        \ell(j):=&2\lambda\bra{\Psi,j}\hn\ket{\Psi,j}=-2\eval{\partial_{\Delta}\log(\bra{\Psi,j}q^{2\Delta \hat{n}}\ket{\Psi,j})}_{\Delta=0}\\
        =&-\qty(\psi_{q^2}\qty(\frac{1}{2}-j)+\psi_{q^2}\qty(\frac{1}{2}+j)+2\log(1-q^2))~,
    \end{aligned}
\end{equation}
with $\psi_q(z):=\partial_z\log(\Gamma_q(z))$ the q-Digamma function. After regularization in the last term, the result (\ref{eq:ell j}) for $q\rightarrow1$ agrees with a BPS wormhole length in $\mathcal{N}=2$ JT gravity \cite{Lin:2022zxd} (69).

\section{Relational Holographic Perspective on Time(less) Evolution}\label{sec:Relational Holo}
In this section, we propose a procedure to treat dressed bulk observables (with or without time dependence) in SUSY theories using a PW-inspired reduction map with respect to reference clock whose QRF orientations (i.e.~the parametrization of its state) are the U$(1)$ R-charge and the boundary time. We specialize most of the analysis to the $\mathcal{N}=2$ DSSYK and its putative bulk dual, although the general formalism developed here is independent of the specific $\mathcal{N}=2$ holographic system.

\paragraph{Outline}In Sec.~\ref{ssec:Bulk super chord} we describe the relational bulk interpretation of the boundary Hilbert space in terms of states in the bulk kinematical and physical Hilbert space. In Sec.~\ref{ssec:relational observables} we describe gravitationally dressed (relational) observables, including those with trivial evolution in the BPS sector. We illustrate the arguments by defining a dressed wormhole length operator from operators acting on the super-chord Hilbert space. This related to the BPS spread complexity proposal in Sec.~\ref{sec:BPS wormholes}.

\subsection{Relational Bulk Interpretation from Super-Chord Hilbert Space}\label{ssec:Bulk super chord}
In the following, we search for a bulk interpretation of physical states in the super-chord Hilbert space (based on the works \cite{Lin:2022rbf,Boruch:2023bte}) to define bulk relational observables through gravitational dressing to the boundary location of the $\mathcal{N}=2$ DSSYK.

We consider arbitrary states within the boundary theory that evolve (or do not evolve in the case of exactly zero-energy states) according to a Schrödinger equation
\begin{equation}\label{eq:Schrödinger}
\rmi\partial_t\ket{\phi}=\hH\ket{\phi}~,
\end{equation}
where $t\in\mathbb{R}$. This can be straightforwardly generalized to complex time used in the HH preparation of state with finite temperatures (e.g.~\cite{Erdmenger:2023wjg,Aguilar-Gutierrez:2025hty,Blommaert:2025eps}); however, for notational simplicity in the relational analysis, that would otherwise contain thermal ensembles, and adopt a specific state preparation, we use real time in this section, and we move to complex-valued times until Sec.~\ref{sec:non BPS N2}.

In the following, we assume that the bulk interpretation of the Schrödinger equation in the boundary theory is the Wheeler-DeWitt \cite{DeWitt:1967yk,Wheeler:1968iap} (WDW) equation (which can be justified e.g.~\cite{Blommaert:2025eps})\footnote{This relation has been found between bosonic DSSYK and sine-dilaton gravity \cite{Blommaert:2025eps}; while there are additional bulk constraints both in the bosonic and SUSY cases, they do not play a role in this discussion.}
\begin{equation}\label{eq:WDW}
\hH_{\rm WDW}\ket{\psi}:=(\hH_{\rm bdry}-\hH_{\rm bulk})\ket{\psi}=0~,\quad \forall\ket{\psi}\in\mathcal{H}_{\rm phys}~,
\end{equation}
where $\mathcal{H}_{\rm phys}$ is the physical bulk Hilbert space, $\hH_{\rm bdry}$ the boundary Hamiltonian (corresponding to the generator of time flow $\rmi\partial_t$), and $\hH_{\rm bulk}$ the Arnowitt–Deser–Misner (ADM) Hamiltonian (corresponding to $\hH$ in the boundary theory). $\mathcal{H}_{\rm phys}$ is constructed by implementing all the constraints on the kinematical Hilbert space ($\mathcal{H}_{\rm kin}$), including \eqref{eq:WDW} and any others that lead to the Hilbert space isomorphism with the boundary theory, such as non-perturbative ones in the genus expansion of the gravitational path integral\footnote{I thank Gonçalo Araujo-Regado for discussions about this.}, corresponding to finite $N$ effects in the boundary theory.

We define $\mathcal{H}_{\rm kin}$ as a tensor product Hilbert space of states where the $\hH_{\rm bdry}$ and $\hH_{\rm bulk}$ operators act separately, denoted the reference $\mathcal{H}_{\rm R}$ and system $\mathcal{H}_{\rm S}$ bulk Hilbert space respectively. To generate the reference state, we propose to incorporate supercharges in the clock state of the QRF constructions \cite{Hoehn:2019fsy} as
\begin{equation}\label{eq:ref clock state}
    \mathcal{H}_{\rm R}:=\qty{\ket{t,j}}_{t\in\mathbb{R},~j\in\mathbb{Z}}~.
\end{equation}
Here, the time and R-charge dependent clock state generalizes the bosonic construction in \cite{Hoehn:2019fsy}:\footnote{In more general SUSY theories that in this setting, the clock state $\ket{t,j}$ can be either bosonic (which we refer to as a ``\emph{referon}'' or ``\emph{framon}'') or fermionic (``\emph{referino}'' or ``\emph{framino}'') \cite{DKS}.}
\begin{equation}\label{eq:clock state}
    \ket{t,j}:=\rme^{-\rmi t\hH_{\rm bdry}}\ket{\Omega,j}~,
\end{equation}
where $\ket{\Omega,j}$ is the maximally entangled state for a fixed R-charge $j$ (while the global one corresponds to $\sum_{j=-\infty}^{\infty}\ket{\Omega,j}$), and $\hH_{\rm bdry}$ acts as the chord Hamiltonian \eqref{eq:H N2 antisymmetric form}. The inner product in $\mathcal{H}_R$ then becomes
\begin{equation}\label{eq:function chi}
    \bra{t,j}\ket{t',j'}=\chi(t-t')\delta_{jj'}~.
\end{equation}
where $\chi(t-t'):=\bra{\Omega,j}\rme^{-\rmi(t-t')\hH_{\rm bdry}}\ket{\Omega,j}$ is a analytic continuation of the SUSY partition \eqref{eq:full partition function of states} where $\beta\rightarrow \rmi(t-t')$; and the factor $\delta_{jj'}$ follows from the definition of the $j$-states \eqref{eq:dif operrators J}. In contrast, there are simplifications in the bosonic case where one can find closed form expressions depending on the spectrum range (see e.g.~(2.7) in \cite{DeVuyst:2024uvd}).

Meanwhile, $\mathcal{H}_{\rm S}$ is defined from the Hilbert space isomorphism to \eqref{eq:physical state N2 jR} prior to implementing \eqref{eq:WDW} as
\begin{equation}\label{eq:system}
\mathcal{H}_{\rm S}:=\qty{\ket{\Omega,j},~\ket{L,XO,j},~\ket{L,OX,j},~\ket{L,XX,j},~\ket{L,OO,j}}_{L\in\mathbb{R},~j\in\mathbb{Z}}~.
\end{equation}
The above definition for the bulk system Hilbert space is an extension of $\mathcal{H}_{\rm super-chord}$ \eqref{eq:physical state N2 jR} where states are labeled by $L\in\mathbb{R}$ (recovering \eqref{eq:physical state N2 jR} when $L\in\mathbb{N}$) prior to imposing both gauge and physical constraints in the bulk, respectively corresponding to time isomorphisms and the momentum shift symmetry (MSS) in sine dilaton gravity \cite{Blommaert:2024whf}, which discretizes the parameter $L$ to take non-negative integer values. However, we stress that the specific definition of \eqref {eq:system} is meant to illustrate how the PW procedure works in the kinematical Hilbert space of the bulk theory dual to $\mathcal{N}=2$ DSSYK (which is assumed to have $L\in\mathbb{R}$ states instead of $n\in\mathbb{N}$ before implementing physical constraints as in sine dilaton gravity \cite{Blommaert:2024whf}). If there were additional states with respect to \eqref{eq:system} in the bulk theory analysis, then one must include the corresponding additional constraints reducing the Hilbert space to be isomorphic to $\mathcal{H}_{\rm super-chord}$ in \eqref{eq:physical state N2 jR} by multiplying with additional projectors (with appropriate operator ordering; see \eqref{eq:equiv class} below) in the physical state equivalence class.\footnote{It would be interesting to verify this explicitly, however, it is outside the scope of this manuscript.} We also adopt the inner product for defining \eqref{eq:system}
\begin{equation}\label{eq:new chord inner}
    \bra{L,AB,j}\ket{L',CD,j'}:=\mathcal{N}_{AB,CD}(L,j)~\delta{(L-L')}\delta_{jj'}~.
\end{equation}
Here $A,~B,~C,~D\in\qty{X,O,\Omega}$ and $\mathcal{N}_{AB,CD}(L,j)$ is a normalization constant; specifically those in \eqref{eq:inner product super chord} for states in the physical Hilbert space, i.e.~after imposing the WDW and MSS constraints, where $L$ is replaced by $n\in\mathbb{N}$. 

Therefore, the kinematical space can be expressed as $\mathcal{H}_{\rm kin}=\mathcal{H}_{\rm S}\otimes\mathcal{H}_{\rm R}$. We then build physical states from the kinematical ones as
\begin{equation}\label{eq:physical state}
    \ket{\psi}_{\rm phys}=\int_{\mathbb{R}}\rmd L\sum_{j,\rm \qty{A,B}}\Psi_{AB}(L,j,j')\ket{L,AB,j}\otimes\ket{t,j'}~,
\end{equation}
where $\Psi_{AB}(L,j,j')$ are constant coefficients (with $A$, $B$ in the same notation as \eqref{eq:new chord inner}) which depend on the bulk constraints to recover a physical state; while $t$ and $j'$ represent the clock readings. One may allow the clock state to be in the same U(1)$_R$ symmetry sector as the system state $\ket{L,AB,j}$ in \eqref{eq:physical state} by simply setting $j'=j$. 

Furthermore, $\ket{\psi}_{\rm phys}$ is by definition annihilated by the constraint $\hH_{\rm WDW}=\hH_{\rm bdry}-\hH_{\rm bulk}$. Using the definition of time state \eqref{eq:clock state} one has that
\begin{equation}\label{eq:time translation constraint}
    \rme^{-\rmi\xi\hH_{\rm bdry}}\ket{t,j}=\ket{t+\xi,j}~,
\end{equation}
and we stress $\hH_{\rm bulk}\ket{n,AB,j}=\hH\ket{n,AB,j}$ for the physical chord states (i.e.~after imposing constraints) as displayed in e.g.~\eqref{eq:H N2 antisymmetric form}, but not for generic states in $\mathcal{H}_{\rm S}$.

Next, we construct equivalence classes of physical states under the bulk constraints using first a group averaging projector (for the time isomorphisms in the bulk) to gauge-invariant states (similar to the literature on QRFs \cite{DeVuyst:2024pop}); and to recover non-divergent bulk physical states, we implement the MSS projector in \cite{Blommaert:2024whf}
\begin{subequations}
    \begin{align}
    &\tilde{\ket{\Psi}}:=\hat{\Pi}_{\rm mss}\hat{\Pi}_{\rm phys}\ket{\psi}~,\quad \ket{\psi}\in\mathcal{H}_{\rm kin}~,\label{eq:equiv class}\\
    &\hat{\Pi}_{\rm phys}:=\int_{\mathbb{R}}\rmd \xi~\rme^{-\rmi\hH_{\rm WDW}\xi}=2\pi\delta(\hat{H}_{\rm WDW})~,\label{eq:curcial projector}\\
    &\hat{\Pi}_{\rm MSS}:=\frac{\prod_{k=-\infty}^\infty\rme^{2\rmi k \hat{L}}}{\prod_{k=-\infty}^\infty1}~,\label{eq:MSS projector}
\end{align}
\end{subequations}
where $\hat{L}\ket{L,AB,j}:=L\ket{L,AB,j}$. Note that the operator ordering is important since $\hat{L}$ and $\hH_{\rm WDW}$ do not commute; one should first perform the group averaging treat the time isomorphism in the physical bulk states, and then implement the MSS that leads to finite norm states. The associated inner products within the equivalence classes are:
\begin{equation}
    \bra{\tilde{\Psi}_1}\ket{\tilde{\Psi}_2}:=\bra{\psi_1}\Pi_{\rm phys}\Pi_{\rm MSS}\Pi_{\rm phys}\ket{\psi_2}~,\quad\ket{\psi_{1,2}}\in\mathcal{H}_{\rm kin}~.
\end{equation}
Next, we build relational states within $\mathcal{H}_{\rm phys}$ by defining a SUSY PW reduction map
\begin{equation}\label{eq:reduction}
    \mathcal{R}(\xi,j):=\mathbb{1}_{\rm S}\otimes\bra{t=\xi,j}~.
\end{equation}
One may add a sum over $j$ in \eqref{eq:reduction} if we were to consider clock superpositions with different R-charges. 

The purpose of the reduction map \eqref{eq:reduction} is to fix the clock readings for conditional states (relative to the frame $R$) defined from \eqref{eq:equiv class} as
\begin{equation}\label{eq:perspective fixed}
    \ket{\psi_{|R}(\xi,j)}:=\mathcal{R}(\xi,j)\tilde{\ket{\Psi}}~,
\end{equation}
which automatically satisfies the Schrödinger equation \eqref{eq:Schrödinger} with $t\rightarrow\xi$ as one can easily verify (see e.g.~(2.36) \cite{DeVuyst:2024uvd}).

To see this, we consider the most general state with a fixed reference clock (in the bulk's boundary) reading at $t=\xi_0$ and in the same U$(1)_{\rm R}$ symmetry sector as the bulk system, given by
\begin{equation}\label{eq:kin state}
    \ket{\psi_{\rm kin}(\xi_0)}:=\int_{\mathbb{R}}\rmd L\sum_{j,\qty{AB}}\Psi_{AB}(L,j)\ket{L,AB,j}\otimes\ket{t=\xi_0,j}~.
\end{equation}
We find that the reduction map \eqref{eq:reduction} with $\ket{\psi}= \ket{\psi_{\rm kin}(\xi_0)}$ \eqref{eq:kin state} in \eqref{eq:equiv class} generates a perspective-fixed evolved state
\begin{equation}\label{eq:perspective reduced state}
\begin{aligned}
    \ket{\psi_{|R}(\xi,j)}&=\qty(\int\rmd\eta~\chi(\eta+\xi-\xi_0)~\rme^{\rmi\hH\eta})\sum_{n}\Psi_{AB}(n,j)\ket{n,AB,j}~,
\end{aligned}
\end{equation}
where we applied \eqref{eq:time translation constraint} after projecting the states \eqref{eq:equiv class}; then we relabeled $L\rightarrow n$; and we used
\begin{equation}\label{eq:below eq}
\hH_{\rm bulk}\ket{n,AB,j}=\hH_{\rm bdry}\ket{n,AB,j}    ~,
\end{equation}
for bulk states obeying the WDW constraint in the last equality. The integral in the parenthesis in \eqref{eq:perspective reduced state} may be evaluated from the boundary perspective where $\hH_{\rm bdry}=\hH$ the $\mathcal{N}=2$ DSSYK Hamiltonian, which means that $\chi(t-t')$ just below \eqref{eq:function chi} is given by the partition function \eqref{eq:full partition function of states} with analytically continuation $\beta\rightarrow\rmi(t-t')$, which does not have a closed form. In particular, notice from \eqref{eq:perspective reduced state} that we recover the BPS HH state $\ket{\Psi,j}$ \eqref{eq:BPS HH state j} by considering the symmetry sector satisfying the supercharge constraints \eqref{eq:supercharge constraints} with $\Psi_{XO}(n,j)=\alpha_n(j_R)$, $\Psi_{OX}(n,j)=\beta_n(j_R)$ in \eqref{eq:alpha bbeta n}, and $\Psi_{AB}(n,j)=0$ otherwise, as well as $\xi=\xi_0$.

\paragraph{Summary:} The above relational framework can be used to formulate the bulk theory dual to the $\mathcal{N}=2$ DSSYK with an explicit observer (the spacetime boundary) probing the bulk interior in a gauge-invariant matter, which may evolve with different parameters (such as the R-charge) besides boundary time. This procedure is based on the bulk kinematical Hilbert space, whose PW reduction generates the physical Hilbert space isomorphic to the super-chord Hilbert space after imposing the corresponding projectors.

\subsection{Time(less) Evolving Relational Observables}\label{ssec:relational observables}
We now implement the previous relations to describe dressed observables with respect to the clock $R$ (the asymptotic boundary) with an R-charge $j$ that makes a time reading at $t=\xi$. The clock dressing is described by the \emph{G-twirl} \cite{Hoehn:2019fsy} (i.e.~incoherent group averaging \cite{DeVuyst:2024uvd})
\begin{equation}\label{eq:Gtwirl}
    \hat{\mathcal{O}}_R^{(t=\xi,j)}(\hat{a}):=\prod_{k=-\infty}^\infty\int_{\mathbb{R}} \rmd\eta~\rme^{-\rmi\hH_{\rm WDW}\eta}\qty(\hat{a}\otimes\ket{t=\xi,j}\bra{t=\xi,j})\rme^{\rmi\hH_{\rm WDW}\eta}~,
\end{equation}
where $\hat{a}$ is an undressed operator in the algebra of the bulk system which is made gauge-invariant through the group averaging with the constraint \eqref{eq:WDW} implemented in \eqref{eq:Gtwirl}. {{The G-twirl represents averaging over the group generated by the constraint (i.e.~$\hH_{\rm WDW}$ in our case) for observables $\hat{a}$ in the kinematical algebra of operators, with a
projection onto the ``clock'' reading of the QRF $\ket{t=\xi,j}$. The bosonic analogue of \eqref{eq:Gtwirl} was introduced for temporal QRFs by \cite{Hoehn:2019fsy} as an extension similar techniques for spatial QRFs without constraints (e.g.~\cite{Bartlett:2006tzx}), in order to build techniques allowing to define relational observables from the perspective neutral framework.}}

One can check that by construction the G-twirl \eqref{eq:Gtwirl} is related to the equivalence class of states \eqref{eq:equiv class} and the perspective-fixed Schrödinger states \eqref{eq:perspective fixed} through
\begin{equation}\label{eq:GT int}
    \bra{\tilde{\Psi}}\hat{O}^{(t=\xi,j)}_{R}(\hat{a})\ket{\tilde{\Psi}}=\bra{\psi_{|R}(t=\xi,j)} \hat{a}\ket{\psi_{|R}(t=\xi,j)}~.
\end{equation}
We now study an example to simplify the above G-twirl by choosing a given element of the bulk dual to super-chord algebra. Let us consider for instance the total bulk wormhole length operator $\hat{L}$ (defined below \eqref{eq:MSS projector}) dual to the total chord number $\hat{n}:=\hat{n}_{X}+\hat{n}_{O}$ in the $\mathcal{N}=2$ DSSYK \cite{Boruch:2023bte} (at least in the semiclassical limit) and the BPS symmetry sector within the super-chord Hilbert space, $\ket{\psi_{|_{\rm R}}}=\ket{\Psi,j}$ \eqref{eq:BPS HH state j} described from the bulk perspective just below \eqref{eq:below eq}. While the wormhole length operator $\hat{L}$ dual to the total chord number by definition acts on all the kinematical bulk states instead of only those obeying the constraints; nevertheless, it can be used to evaluate expectation values of the bulk relational observables in \eqref{eq:GT int}, such as
\begin{equation}\label{eq:two point secretly}
    \bra{\tilde{\Psi}}\hat{\mathcal{O}}_R^{(j)}\qty(\rme^{-\Delta\hbar\hat{L}})\ket{\tilde{\Psi}}=\bra{\Psi,j}\rme^{-\Delta\hbar\hat{L}}\ket{\Psi,j}~,
\end{equation}
where $\hbar$ and $\Delta$ are constants (the latter corresponds to the conformal dimension of the matter operators so that \eqref{eq:two point secretly} reproduces a two-point function in $\mathcal{N}=2$ DSSYK \cite{Berkooz:2020xne,Boruch:2023bte}), and we suppressed the trivial $\xi$ index. The expectation value \eqref{eq:two point secretly} can be computed, and the boundary side interpretation in our construction above corresponds to a matter two-point function in \cite{Boruch:2023bte} (4.5)). Therefore, while $\hat{L}$ is not by itself an operator in the physical operator algebra (since it acts on all kinematical states), it can still be used to evaluate dressed observables that may or may not evolve trivially. In the following section we study in more detail this example in terms of complexity growth for BPS and non-BPS states.

\section{BPS spread complexity in \texorpdfstring{$\mathcal{N}=2$}{} Double-Scaled SYK}\label{sec:BPS wormholes}
In this section, we formulate a new definition of spread complexity which is well-adapted to describe zero-energy BPS states, where the complexity growth is determined by the R-charge parameterization. {{The original definition of spread complexity \cite{Balasubramanian:2022tpr} is stated in Sec.~\ref{sec:non BPS N2}.}} We begin with Sec.~\ref{ssec:R charge evolution} formulating a Lanczos algorithm and a corresponding Krylov basis for the BPS HH state. The method is based on an effective Hamiltonian determined by the BPS HH state \eqref{eq:BPS HH state j}. In Sec.~\ref{ssec:BPS Krylov proposal}, we describe our proposal for the spread complexity of BPS states based on the previously derived Krylov basis, which we analyze. The proposed notion of spread complexity exactly matches the expectation value total chord number in the BPS HH state \cite{Boruch:2023bte}, as well as a semiclassical two-sided AdS wormhole length in $\mathcal{N}=2$ JT supergravity \cite{Lin:2022zxd}.

\subsection{Krylov Basis from R-Charge ``Evolution'' in the BPS state}\label{ssec:R charge evolution}
We start from the BPS HH state \eqref{eq:BPS HH state j}. We notice that from \eqref{eq:alpha bbeta n} that the coefficients of the state, determined by the supercharge constraints, can be expressed as
\begin{equation}\label{eq:decomposition alpha betaa n}
    \alpha_n:=\bra{j_R}\ket{B_n}~,\quad\beta_n:=\bra{-j_R}\ket{B_n}~,
\end{equation}
where $\ket{B_n}$ is an auxiliary chord number basis, and $\ket{j_R}$ also denotes an auxiliary R-charge basis such that the inner product determines the BPS state coefficients \eqref{eq:decomposition alpha betaa n}. We can then express both relations in \eqref{eq:coefficients alpha beta} in terms of an effective Hamiltonian\footnote{I thank Jiuci Xu for pointing out there should be an effective Hamiltonian in this construction.}, namely\footnote{Note that although the effective Hamiltonian might appear to be non-Hermitian, this depends on the choice of inner product. As in \cite{Aguilar-Gutierrez:2025hty} the Hermicity of the effective Hamiltonian follows from the commutation relations of the operators in \eqref{eq:n hat Bn} and the Hermitian conjugate operation of the corresponding $*$-algebra \cite{Lin:2023trc,Xu:2024hoc}.}
\begin{align}\label{eq:alpha beta basis}
\hH'_{\rm eff}\ket{B_n}=\qty(\qty(q^3-q^{2\hn-1})\rme^{\rmi\hP}+\rme^{-\rmi\hP})\ket{B_n}~,
\end{align} 
where
\begin{subequations}
\begin{align}
&\hH'_{\rm eff}\ket{j_R}:=\qty(q^{-j_R-2}+q^{j_R-1})\ket{j_R}~,\label{eq:effective H R-charge}\\
&\hn\ket{B_n}:=n\ket{B_n}~,\quad \rme^{\pm\rmi\hP}\ket{B_n}:=\ket{B_{n\pm1}}~.\label{eq:n hat Bn}
\end{align}
\end{subequations}
Note that the overall normalization of the wavefunction does not play a role in the deriving \eqref{eq:alpha beta basis}.

We now look for a orthonormal basis where the effective Hamiltonian remains tridiagonal by applying a canonical transformation, corresponding to a change of the $\ket{B_n}$ basis by
\begin{equation}
    \rme^{-\rmi \hP}\rightarrow\rme^{-\rmi \hP}\sqrt{q^3-q^{2\hn-1}}~,\quad \rme^{\rmi \hP}\rightarrow\qty(q^3-q^{2\hn-1})^{-1/2}\rme^{\rmi \hP}~,
\end{equation}
such that we can recognize the Krylov basis more easily.

This means that \eqref{eq:alpha beta basis} can be written in terms of an effective Hamiltonian $\hH'_{\rm eff}\rightarrow\hH_{\rm eff}$ obeying a recursion relation of the form
    \begin{align}
    \hH_{\rm eff}\ket{K_n}=b_{n+1}\ket{K_{n+1}}+b_n\ket{K_{n-1}}~,    
    \end{align}
where the initial state in the Lanczos algorithm for the effective Hamiltonian is chosen as $\ket{K_2}=\ket{B_0}$ with\footnote{Alternatively, one can take $\ket{K_0}=\ket{B_0}$ as the initial state in the algorithm; however, the index $n$ in \eqref{eq:eigen val problem N2} should be shifted $n\rightarrow n+2$ to recover agreement with the requirement $b_0=0$ for the initial state in the Lanczos algorithm \cite{Balasubramanian:2022tpr}. We thank Jiuci Xu for related remarks about this in a previous draft.}
\begin{equation}\label{eq:eigen val problem N2}
    b_{n\geq2}=\sqrt{q^{3}-q^{2n-1}}~,\quad     q\in[0,1)~.
\end{equation}
Note that both $\ket{B_{n\geq0}}$ and $\ket{K_{n\geq 2}}$ form a complete basis
\begin{equation}\label{eq:resolution ide}
    \mathbb{1}=\sum_{n=0}^\infty\ket{B_n}\bra{B_n}=\sum_{n=2}^\infty\ket{K_n}\bra{K_n}~.
\end{equation}
Now that we recovered a Krylov basis given a reference state $\ket{K_2}$ and the effective Hamiltonian derived by the supercharge constraints \eqref{eq:supercharge constraints} in the BPS state, it is natural to ask
\begin{quote}
    \emph{What is the corresponding semiclassical Krylov complexity of the BPS state? How is it related to the wormhole length \eqref{eq:ell j}?}
\end{quote}
We study this {{and answer it positively}} in the next subsection. In App.~\ref{app:alternative} we provide an alternative approach, where we instead apply the original definition of spread complexity using the effective Hamiltonian \eqref{eq:effective H R-charge} to evolve a reference state $\ket{B_0}$. Since the resulting measure does not encode information about the states $\ket{n,XO,j}$, $\ket{n,OX,j_R}$, but rather about $\ket{B_n}$, it does not result in the expectation value of the chord number in the BPS state \eqref{eq:BPS HH state j}. For this reason, we define a more appropriate measure accounting for $\ket{n,XO,j}$, $\ket{n,OX,j}$ and the coefficients $\alpha_n$, $\beta_n$.

\subsection{A Proposal for BPS Spread Complexity}\label{ssec:BPS Krylov proposal}
We propose that to associate an extended notion of Krylov complexity to a BPS wormhole \cite{Lin:2022zxd,Lin:2022rzw}, one should define a map, denoted $\hat{\mathcal{L}}$, that takes the R-charge coefficients (defining the Krylov basis) to states in a doubled Hilbert space\footnote{This is equivalent to the Choi–Jamiołkowski isomorphism \cite{jamiolkowski1972linear,choi1975completely} used to evaluate e.g.~Krylov operator complexity \cite{Parker:2018yvk}.}
\begin{equation}\label{eq:choi}
\begin{aligned}
\hat{\mathcal{L}}:&~\bra{\pm j_R}\ket{B_n}\rightarrow|\pm j_R,B_n)~,\\
\hat{\mathcal{L}}^\dagger:&~\bra{B_n}\ket{\pm j_R}\rightarrow(B_n,\pm j_R|~,
\end{aligned}
\end{equation}
where in this notation $|a,b):=\ket{a}\otimes\ket{b}$. Using the above Choi–Jamiołkowski isomorphism, we can represent the BPS state \eqref{eq:BPS HH state j} as
\begin{equation}
\hat{\mathcal{L}}:~\ket{\Psi,j}\rightarrow|\Psi_j):=\sum_n\qty[|B_n,j_R)\ket{n,XO,j}+|B_n,-j_R)\ket{n,OX,j}]~,
\end{equation}
and we define the Krylov complexity operator for BPS state in terms of the doubled Krylov states as\footnote{The second relation follows from the fact that $\ket{B_n}=B_n\ket{K_{n+2}}$ while $\bra{B_n}=1/B_n\bra{K_n}$ (so that $\ket{K_{n+2}}\bra{K_{n+2}}=\ket{B_n}\bra{B_n}$) by construction, which allows the resolution of the identity in either basis \eqref{eq:resolution ide}.}
\begin{equation}\label{eq: def C_d}
\hat{\mathcal{C}}_d:=\sum_{n=2}^\infty(n-2)|K_n,K_n)(K_n,K_n|=\sum_{n=0}^\infty n|B_n,B_n)(B_n,B_n|~,
\end{equation}
where the subindex $d$ denotes doubled. Using the operators above, we define the (unnormalized) BPS spread complexity for the reference state \eqref{eq:BPS HH state j} as
\begin{equation}\label{eq:BPS wormhole length}
\begin{aligned}
\mathcal{C}_d:=(\Psi_j|\hat{\mathcal{C}}_d|\Psi_j)=\sum_{n=0}^\infty &n\bigg[\abs{\alpha_n}^2\norm{n,XO,j}^2+\abs{\beta_n}^2\norm{n,OX,j}^2\\
+&\alpha_n\beta_n^*\bra{n,OX,j}\ket{n,XO,j}+\alpha^*_n\beta_n\bra{n,XO,j}\ket{n,OX,j}\bigg]~,
\end{aligned}
\end{equation}
where in the last relation we carried out the inner product within the doubled Hilbert space, where e.g.~$(B_m,B_m|j_R,B_n)=\alpha_n$ together with the chord inner product for elements $\bra{n,AB,j}\ket{m,CD,j}$ (with $A$, $B$, $C$, $D\in\qty{X,O,\Omega}$) in \eqref{eq:inner product super chord}, and similarly for the other elements.

Thus, the proposal for BPS spread complexity \eqref{eq:BPS wormhole length} reproduces the total chord number operator in the BPS HH state \cite{Boruch:2023bte} (which, again, considers the normalization of states in \cite{Berkooz:2020xne}) without using semiclassical approximations. Then, the advantage of the proposal \eqref{eq: def C_d} for spread complexity of BPS states is that it reproduces a bulk wormhole answer, extending the purely bosonic results in the literature so far. Note also that in defining \eqref{eq:BPS wormhole length} we selected a basis of states that \emph{tridiagonalize} the effective Hamiltonian \eqref{eq:effective H R-charge} determined from the supercharge constraints \eqref{eq:supercharge constraints}. Thus, the proposed measure of spread complexity is basis dependent, similar to the original proposal by \cite{Balasubramanian:2022tpr}, and it is determined by a Lanczos algorithm. We provide further comments about extensions of the proposal in Sec.~\ref{ssec:outlook}.

\paragraph{Comparison of the proposal with the literature}
{In this section, we employed the R-charge $j$ (related to $j_R$ through \eqref{eq;def jR}) as a measure of evolution of the BPS state \eqref{eq:BPS HH state j} and the corresponding BPS wormhole length. We stress that the notion of spread complexity that we defined is different from other proposals in the literature. For instance, in symmetry resolved spread complexity \cite{Caputa:2025ozd} one would use the R-charge to separate symmetry sectors in the time evolution of spread complexity; while here the R-charge itself determines the evolution. Meanwhile, in contrast to generalized Krylov complexity \cite{FarajiAstaneh:2025thi}, we do use a generator of evolution in $j$, since here it corresponds to physical parameter determining the U(1) symmetry sector here, we instead work at the level of the constraints of the supercharges on the coefficients of the BPS HH state. It would be interesting to extend this comparison}.

\section{Non-BPS Wormhole Lengths from the \texorpdfstring{$\mathcal{N}=2$}{} Double-Scaled SYK}\label{sec:non BPS N2}
In this section, we study the original notion of spread complexity \cite{Balasubramanian:2022tpr} {{(see \eqref{eq:Krylov exp} below)}} for different states with non-trivial boundary time dependence in the $\mathcal{N}=2$ DSSYK model. In Sec.~\ref{ssec:bosonic subspaces} we work on this problem for states within the orthogonal bosonic subspaces of Sec.~\ref{ssec:aux system N2}. In Sec.~\ref{ssec:TFD wormhole} we study the Krylov space and spread complexity of the HH state built from the maximal entangled state at fixed R-charge.
In all cases, we recover the same semiclassical evolution of spread complexity as in the purely bosonic case \cite{Rabinovici:2023yex}. However, quantum corrections do contain this information in the latter case (Sec.~\ref{ssec:TFD wormhole}).

\subsection{Bosonic Orthogonal Subspaces}\label{ssec:bosonic subspaces}
In this subsection, we study the Krylov space spanned by the bosonic states \eqref{eq:ortho bosonic basis} of the zero particle super-chord Hilbert space, and the spread complexity of the corresponding HH state within each subspace.  

\paragraph{Representation for the Hamiltonian}
First note that the basis in \eqref{eq:ortho bosonic basis} corresponds to orthogonal subspaces (i.e.~$\bra{\bar H_n}\ket{H_m}=0$ $\forall m,~n$) of the bosonic sector of the spectrum (which we can denote $\mathcal{B}$ and $\bar{\mathcal{B}}$ as in \cite{Berkooz:2020xne}). This allows us to identify an operator relation for the Hamiltonian acting on the set of states $\mathcal{B}\cup\bar{\mathcal{B}}$,\footnote{Note that $\mathcal{B}\cup\bar{\mathcal{B}}$ is not necessarily spanned by the set of states build from the Hamiltonian acting on the zero-chord state $\qty{\ket{\Omega,j},~\hH\ket{\Omega,j},~\hH^2\ket{\Omega,j},\dots}$.}
\begin{equation}\label{eq:H N2 operator}
    \hH=q^{-1/2}k\qty(\rme^{-\rmi \hP}+\rme^{\rmi \hP}\qty(1-q^{2\hn})+\qty(q^{-\hs+1/2}+q^{\hs-1/2}))~,
\end{equation}
where
\begin{subequations}\label{eq:H Hbar}
  \begin{align}
    &\rme^{\pm\rmi \hP}\ket{H_n}=\ket{H_{n\pm1}}~,\quad &&\hat{n}\ket{H_n}=n\ket{H_n}~,\quad &&\hat{j}_R\ket{H_n}=j_R\ket{H_n}~,\label{eq:Hn basis}\\
    &\rme^{\pm\rmi \hP}\ket{\bar H_n}=\ket{\bar H_{n\pm1}}~,\quad &&\hat{n}\ket{\bar H_n}=n\ket{\bar H_n}~,\quad &&\hat{j}_R\ket{\bar H_n}=-j_R\ket{\bar H_n}~.
\end{align}  
\end{subequations}
We emphasize that \eqref{eq:H N2 operator} does not connect $\ket{\Omega,j}$ with $\mathcal{B}\cup\bar{\mathcal{B}}$, since it acts in a different way on $\ket{\Omega,j}$ (as seen in \eqref{eq:H N2 antisymmetric form_a}); thus, one needs to carry out the Lanczos algorithm for this separately (see Sec.~\ref{ssec:TFD wormhole}).

\paragraph{Krylov Basis and spread complexity}
Performing the canonical transformation
\begin{equation}
    \rme^{-\rmi \hP}\rightarrow\sqrt{1-q^{2\hat{n}}}\rme^{-\rmi \hP}~,\quad \rme^{\rmi \hP}\rightarrow\rme^{\rmi \hP}\qty(1-q^{2\hat{n}})^{-1/2}~,
\end{equation}
(\ref{eq:H N2 operator}) transforms into a symmetric form
\begin{equation}\label{eq:H N2 Krylov basis}
    \hH=k~q^{-1/2}\qty(\sqrt{1-q^{2\hn}}\rme^{-\rmi \hP}+\rme^{\rmi \hP}\sqrt{1-q^{2\hn}}+2\cosh\qty(\lambda\qty(\hs-\frac{1}{2})))~.
\end{equation}
We now build the Krylov basis associated to \eqref{eq:H N2 Krylov basis} acting on either $\ket{H_0}$ or $\ket{\bar{H}_0}$ as the reference state in the algorithm, which we denote as
\begin{equation}\label{eq:starting Krylov}
    \ket{K_0}=\qty{\ket{H_0},~\ket{\bar{H}_0}}~.
\end{equation}
Note that although there are two reference states, we construct a Krylov basis for each one ($\ket{H_0}$, $\ket{\bar H_0}$), and in either case, the last term in \eqref{eq:H N2 Krylov basis} is an overall constant (in contrast to bosonic DSSYK), i.e.~independent on the index $n$ in the Lanczos algorithm. This is a consequence of $\hH$ acting only on the bosonic states of the model instead of the true ground states (which we study in Sec.~\ref{sec:BPS wormholes}). For this reason, this approach does not probe a symmetric spectrum in this model using $\ket{H_0}$ and $\ket{\bar H_0}$ as reference states for each algorithm; although, this only amounts to an overall shift in the spectrum.

We can subtract the overall constant in the Hamiltonian\footnote{Equivalently, one could keep the overall constants; then, the Krylov basis representation  for (\ref{eq:starting Krylov}), which we denote $\ket{K_n}$ and $\ket{\overline{K}_n}$ corresponding to the different energies, becomes
\begin{subequations}\label{eq:H N2 Krylov form}
\begin{align}
    &\hH\ket{K_n}=b_{n+1}\ket{K_{n+1}}+b_n\ket{K_{n-1}}+a({j_R})\ket{K_n}~,\\
    &\hH\ket{\bar K_n}=b_{n+1}\ket{\bar K_{n+1}}+b_n\ket{\bar K_{n-1}}+a(-{j_R})\ket{\bar K_n}~,
\end{align}
\end{subequations}
where
\begin{equation}
    b_n=k\sqrt{1-q^{2n}}~,\quad a({j_R})=2k\cosh(\lambda({j_R}-1/2))~,
\end{equation}
\begin{equation}
\bra{K_n}\ket{K_m}=\delta_{nm}~,\quad \bra{\bar K_n}\ket{\bar K_m}=\delta_{nm}~,\quad \bra{K_n}\ket{\bar K_m}=0~.    
\end{equation}}
\begin{equation}
    \Delta\hH:=\hH-2k\cosh\qty(\lambda\qty(\hs-\frac{1}{2}))~,
\end{equation}
which has the same Krylov basis of the bosonic DSSYK model without matter chords \cite{Rabinovici:2023yex}, that we denote $\ket{K_n}$, {{which is defined by the recurrence relation}}
\begin{equation}\label{eq:algo eq}
\begin{aligned}
    &\Delta\hH\ket{K_n}=b_{n+1}\ket{K_{n+1}}+b_{n}\ket{K_{n-1}}+a_{n}\ket{K_{n}}~,\\
    &\rme^{\pm\rmi \hP}\ket{K_n}=\ket{K_{n\pm1}}~,\quad \hat{n}\ket{K_n}=n\ket{K_n}~,
\end{aligned}
\end{equation}
{{where $a_n$ and $b_n$ are called the Lanczos coefficients. Explicitly, for Hamiltonian \eqref{eq:algo eq} we have that $a_n=0$ and $b_n=k\sqrt{1-q^{2n}}$.}}

The Krylov basis can then be used to express an evolved state
\begin{equation}\label{eq:initial state N2 orth B}
    \ket{\psi(\tau)}=\rme^{-\tau~\Delta\hH}\ket{K_0}=\sum_{n=0}^\infty \Psi_n(\tau)\ket{K_n}~,
\end{equation}
where $\tau:=\frac{\beta}{2}+\rmi t$ is an analytically continued time {{in the HH preparation of state, where}} $t$ is real time, and $\beta$ corresponds to the inverse temperature, while
\begin{equation}
\begin{aligned}
    \Psi_n(\tau):&=\bra{K_n}\rme^{-\tau~\Delta\hH}\ket{K_0}~,
\end{aligned}
\end{equation}
which obeys the recurrence relation $-\partial_\tau\Psi_n=b_{n+1}\Psi_{n+1}+b_{n}\Psi_{n-1}$.

We can now evaluate the spread complexity with \eqref{eq:initial state N2 orth B} as the reference state {{which, in terms of complex-valued time, is defined as \cite{Erdmenger:2023wjg}}}
\begin{equation}\label{eq:Krylov exp}
    \mathcal{C}(t):=\eval{\frac{\sum_{n=0}^\infty n \abs{\Psi_n(\tau)}^2}{\sum_{n=0}^\infty \abs{\Psi_n(\tau)}^2}}_{\tau=\frac{\beta}{2}+\rmi t}~.
\end{equation}
Due to \eqref{eq:algo eq} this exactly reproduces the same spread complexity for the HH state in the bosonic DSSYK.\footnote{The reader is referred to \cite{Rabinovici:2023yex,Xu:2024gfm,Heller:2024ldz} for numerical approaches $\forall q\in[0,1)$.} In App.~\ref{app:detail ortho eval}, we confirmed this explicitly by working on the $\mathcal{N}=2$ DSSYK model path integral and performing a semiclassical approximation for the chord number, from which we recover (see (\ref{eq:length path integral}, \ref{eq:ell times})):
\begin{equation}\label{eq:complexity ortho}
    \mathcal{C}(t)\eqlambda\frac{2}{\lambda}\log(\frac{\cosh(J\sin\theta ~t)}{\sin\theta})~,
\end{equation}
where we denote the microcanonical temperature as $\beta(\theta)=(\pi-\theta)/(2J\sin\theta)$ (see App.~\ref{app:thermo}), and we choose $k=J/\lambda$ with $J\in\mathbb{R}$.

The result in \eqref{eq:complexity ortho} means that the microstates in $\mathcal{B}\cup\overline{\mathcal{B}}$ in the $\mathcal{N}=2$ DSSYK reproduces the same semiclassical spread complexity and Krylov basis of bosonic DSSYK without incorporating matter chords \cite{Rabinovici:2023yex,Heller:2024ldz}. This is also expected by analyzing wormhole lengths in the dual gravitational description \cite{Fan:2021wsb}. We should note that the Krylov basis of the orthogonal bosonic sector (with $\ket{K_0}$ and $\ket{\bar K_0}$ as the reference) is exactly the same as for the bosonic DSSYK \cite{Ambrosini:2024sre,Heller:2024ldz}, as one would expect. In contrast, we will find that if one studies spread complexity with the HH state built from the zero-chord state of the $\mathcal{N}=2$ model as a reference (see Sec.~\ref{ssec:TFD wormhole}) there are quantum corrections that allow one to differentiate with respect to the bosonic case; and the difference is even sharper for the BPS HH state as a reference (Sec.~\ref{sec:BPS wormholes}).

\subsection{Hartle-Hawking from Maximally Entangled State}\label{ssec:TFD wormhole}
Given that the maximally entangled state for a fixed R-charge $j$ \cite{Berkooz:2020xne}
\begin{equation}
    \ket{K_0}:=\ket{\Omega,j}~,
\end{equation}
is used to defined the partition function of the theory (see App.~\ref{app:thermo} for details) and the corresponding HH state $\rme^{-\beta\hH}\ket{\Omega,j}$ encodes both BPS and non-BPS state contributions (as first observed in \cite{Berkooz:2020xne}), we will now find the Krylov basis associated to it:
\begin{equation}\label{eq:Krylov basis}
    \hH\ket{K_n}=b_{n+1}\ket{K_{n+1}}+b_n\ket{K_{n-1}}+a_n\ket{K_n}~,
\end{equation}
using the fact that the $\mathcal{N}=2$ DSSYK Hamiltonian acting on $\ket{\Omega,j}$ is \eqref{eq:H N2 antisymmetric form_a}.

\paragraph{Building the Krylov basis}
From \eqref{eq:H N2 antisymmetric form}, we know that
\begin{equation}
    \hH\ket{K_0}=\frac{J}{\lambda}\qty(\ket{1,XO,j}+\ket{1,OX,j}+(q^{j_R}+q^{-j_R})\ket{K_0})~,
\end{equation}
 where we adopt as overall scaling of the Hamiltonian
 \begin{equation}\label{eq:parametrization k}
    k:=\frac{J\sqrt{q}}{\lambda}~.
\end{equation}
Here $J\in\mathbb{R}$ is a constant, which does not modify the system, and it allows a straightforward saddle point analysis for the DSSYK path integral.

To deduce the Krylov basis, we consider the orthonormal basis (first appearing in \cite{Berkooz:2020xne})
\begin{equation}\label{eq:Krylov basis N2 Omega}
    \ket{K_n}=\sqrt{\frac{q^n}{2(q^2;q^2)_{n-1}(1-q^n)}}\qty(\ket{n,XO,j}+\ket{n,OX,j})~.
\end{equation}
which is an eigenstate of the total chord number operator. By applying the Hamiltonian to \eqref{eq:Krylov basis N2 Omega} with the rules \eqref{eq:H N2 antisymmetric form}, one recovers in general that:
\begin{equation}\label{eq:Krylov basis more general}
\begin{aligned}
    \hH\ket{K_n}=&b_{n+1}\ket{K_{n+1}}+\frac{q^{-j_R-1}+q^{j_R}+q^{n-j_R}-q^{n-j_R-1}}{\sqrt{2~q^{-n}(q^2;q^2)_{n-1}(1-q^n)}}\ket{n,OX,j}\\
    &+\frac{q^{j_R+n}-q^{n+j_R-1}+q^{j_R-1}+q^{-j_R}}{\sqrt{2~q^{-n}(q^2;q^2)_{n-1}(1-q^n)}}\ket{n,XO,j}+b_n\ket{K_{n-1}}~.
\end{aligned}
\end{equation}
This is very close to the Krylov basis form \eqref{eq:Krylov basis}, albeit not the same for the most general $q\in[0,1)$ and $j\in\mathbb{R}$ $\forall n$. Regardless, it indeed satisfies the Lanczos algorithm exactly up to $n=2$. This means that the Hamiltonian is not tridiagonal in the basis (\ref{eq:Krylov basis N2 Omega}) unless we restrict ourselves to the following cases:
\begin{itemize}
    \item $q\in[0,1)$ and $j=0$,
\begin{subequations}
    \begin{align}
        a_n&=\frac{J}{\lambda}\qty(q^{-1}+1+q^{n}-q^{n-1})~,\\
        b_n&=\frac{J}{\lambda}\sqrt{q^{-1}(1-q^n)(1+q^{n-1})}~.
    \end{align}
\end{subequations}
In order to simplify the subsequent analysis, one needs to further consider $\lambda\rightarrow0$, so we will focus mostly on the case below.
\item $q\rightarrow1^-$ and $j\sim\mathcal{O}(1)$,
    \begin{align}
        a_n&=\frac{2J}{\lambda}~,\quad b_n=\frac{J}{\lambda}\sqrt{1-q^{2n}}~.
    \end{align}
Thus, in the semiclassical limit, where we expect to find a dual classical gravity theory, we recover the bosonic DSSYK results (see e.g.~\cite{Rabinovici:2023yex}). Note that $\ket{K_n}$ in (\ref{eq:Krylov basis N2 Omega}) has norm $\bra{K_n}\ket{K_n}=1~\forall q\in[0,1)$ including the $\lambda\rightarrow0$ limit, so even though $(q^2;q^2)_{n-1}$ diverges when $\lambda\rightarrow0$, the state as a whole does not.
\end{itemize}
We stress that analyzing the above cases allow us to simplify the analysis of the Krylov basis in \eqref{eq:Krylov basis N2 Omega}. Nevertheless, there should exist a more general solution of the Lanczos algorithm for $j\neq 0$. We also explore an alternative (albeit non-Krylov) basis orthogonal to \eqref{eq:Krylov basis N2 Omega} in App.~\ref{app:alternative basis}.

Now, we express the Hamiltonian in the Krylov basis $\qty{\ket{K_n}}$ (\ref{eq:Krylov basis N2 Omega}) for the reference $\ket{K_0}=\ket{\Omega,j}$ as
\begin{equation}\label{eq:Heritian Hamiltonian Omega N2}
\begin{aligned}
    \hH=\frac{J}{\lambda}\bigg(&\sqrt{(1-q^{\hn})(1+q^{\hn-1})}\rme^{-\rmi \hP}+\rme^{\rmi \hP}\sqrt{(1-q^{\hn})(1+q^{\hn-1})}\\
    &+2\qty(\mathbb{1}\cosh\frac{\lambda}{2}+q^{\hn}\sinh\frac{\lambda}{2})\bigg)~.
\end{aligned}
\end{equation}
We then evaluate the spread complexity {{(as defined in \eqref{eq:Krylov exp})}} for the state $\ket{\psi(\tau)}=\rme^{-\tau\hH}\ket{\Omega,j}$, where $\ket{\Omega,j}$ is the reference state in the Lanczos algorithm. The details of the evaluation are relegated to App.~\ref{app:spread zero chord}. The result (see \eqref{eq:spread comple semiclassical}) is
\begin{align}\label{eq:complexity zero chord}
    &\mathcal{C}(t):=\eval{\frac{\sum_nn\abs{\bra{K_n}\ket{\psi(\tau)}}^2}{\bra{\psi(\tau)}\ket{\psi(\tau)}}}_{\tau=\frac{\beta}{2}+\rmi t}\eqlambda\frac{1}{\lambda}\log\frac{\cosh(J\sin\theta~t)}{\sin\theta}~.
\end{align}

\paragraph{Physical Interpretation}As seen from \eqref{eq:complexity zero chord}, we recover the same answer at the semiclassical level as the purely bosonic DSSYK case \cite{Rabinovici:2023yex,Heller:2024ldz} (similar to Sec.~\ref{ssec:bosonic subspaces}). This is expected since the Krylov basis only depends on bosonic states, and not on the R-charge. From a bulk perspective, we expect that the result can be interpreted in a very similar way as the $\mathcal{N}=1$ JT gravity case studied in \cite{Fan:2021wsb} (see below their (5.27)) where the geodesic lengths are identical at late-times as the purely bosonic counterpart. It is only when we include quantum corrections that this is no longer true. We can see this from our construction of the Krylov basis when $j\neq0$ in \eqref{eq:Krylov basis more general}, the Krylov basis and thus the spread complexity is modified as soon as we incorporate the first leading order quantum correction to the semiclassical result (in contrast to Sec.~\ref{ssec:bosonic subspaces}) which is controlled by the R-charge $j$.

\paragraph{Chord number vs spread complexity and bulk wormhole length}
Using the previous definitions, one can see that the $\ket{K_n}$ basis \eqref{eq:Krylov basis N2 Omega} (and similarly for $\ket{L_n}$, which is defined in App.~\ref{app:alternative}) is an eigenstate of the total chord number, i.e.
\begin{equation}
    \hat{N}\ket{K_n}=n\ket{K_n}~.
\end{equation}
Note that each of the basis is orthonormal. However, as we have seen, the $\ket{K_n}$ basis satisfies the Lanczos algorithm only at leading order when $\lambda\rightarrow0$. This means that only in this limit, we have an equality between spread complexity of an state and the expectation value on the same state
\begin{equation}
    \expval{\hat{N}}\eqlambda\mathcal{C}~.
\end{equation}
This also implies that the two-point correlation function \cite{Berkooz:2020xne}
\begin{equation}
    \frac{\bra{\Psi}q^{\Delta\hat{N}}\ket{\Psi}}{\bra{\Psi}\ket{\Psi}}~,
\end{equation}
is the generating function of Krylov complexity only when $\lambda\rightarrow0$. This result shares similarities with Krylov complexity in the bosonic DSSYK model with matter \cite{Aguilar-Gutierrez:2025mxf,Aguilar-Gutierrez:2025pqp,Ambrosini:2024sre}. In contrast, the spread complexity of the HH state as reference state in the bosonic DSSYK model without matter always equals the expectation value of the chord number in the corresponding evolved reference state  \cite{Rabinovici:2023yex,Heller:2024ldz}. We can thus see that, although the states generated by the Lanczos algorithm are bosonic, the fermionic corrections and the presence of R-charge in the theory are still present within Krylov complexity beyond leading order in $\lambda$, as expected also from the fact that $\rme^{-\beta\hH}\ket{\Omega,j}$ encodes information about all the spectrum at a fixed R-charge, including the zero-energy ground state. Thus, the results imply that spread complexity again matches wormhole length for a non-BPS HH state in the semiclassical limit, but there are corrections away from this limit.

\section{Discussion}\label{sec:discussion}
\paragraph{Summary}
In the first part of this work, we formulated  a relational holographic framework to describe bulk dressed operators with or without boundary time evolution, while specializing the arguments to the $\mathcal{N}=2$ DSSYK. This was done by incorporating the R-charge in the clock states within the kinematical bulk Hilbert space and properly treating the bulk constraints in an extension of the PW mechanism. Among the relational observables, we identified the bulk wormhole length dual to the total chord number of the $\mathcal{N}=2$ DSSYK. We emphasize that one can describe the bulk theory relationally in the sense of the PW reduction map \cite{Page:1983uc,delaHamette:2021oex} through the holographic Hilbert space isomorphism to the boundary theory (as well the relation between the bulk WDW/ boundary Schrödinger equations) in terms of an observer (the boundary) and a system (the bulk interior). In contrast, this is not directly possible in the boundary theory (as described at the beginning of Sec.~\ref{sec:intro}). In the particular SUSY setting, this formalism clarifies how dressed observables with or without boundary time evolution, have non-trivial properties according to an observer due to the R-charge dependence.

In the second part, to specialize in natural observables that reveal information about the holographic dictionary, we (i) proposed a new notion of Krylov complexity for BPS states, and (ii) analyzed the original definition of spread complexity \cite{Balasubramanian:2022tpr} for non-BPS states in the $\mathcal{N}=2$ DSSYK model. Concerning (i), the proposal allows to meaningfully associate complexity growth (following a Lanczos algorithm) with respect to the R-charge of BPS states. We showed that this measure exactly reproduces the expectation value of the total chord number in the same reference state, and therefore it also holographically reproduces the wormhole length in $\mathcal{N}=2$ JT supergravity in the semiclassical limit. Meanwhile in (ii), we explored the similarities and differences between spread complexity for non-BPS states in the $\mathcal{N}=2$ model with those in bosonic DSSYK. We showed that despite the existence of different bosonic Krylov basis in the $\mathcal{N}=2$ DSSYK, the spread complexity of an associated non-BPS HH state on those subspaces, at the semiclassical limit reproduces the same answer as in the bosonic DSSYK. Meanwhile, quantum corrections in the Krylov basis encode the differences between the models, such as the R-charge dependence. We also pointed out that from JT supergravity agree with the findings when spread complexity of the reference state is identified with a wormhole length. Thus, our results suggest that one should not give up on quantifying chaos with complexity growth in the zero-energy BPS sectors due to its lack of time evolution; there are other properties, including supercharge constraints, that lead to an emergent Lanczos algorithm associated to the BPS states, and a natural notion of spread complexity with a holographic interpretation. 

We hope that an extension of these ideas can be used to understand other complex systems with trivial time evolution from a relational perspective. We now comment on future research directions.

\subsection{Particle Super-Chord Space \& Entangler Map}\label{ssec:particle space}
Recently, there has been immense progress in understanding the holographic duality of the bosonic DSSYK model by computing correlation functions from an extended chord Hilbert space with matter insertions \cite{Cui:2025sgy,Ambrosini:2024sre,Xu:2024gfm,Aguilar-Gutierrez:2025pqp,Aguilar-Gutierrez:2025mxf,Aguilar-Gutierrez:2025hty,Xu:2024hoc}, which was sparked by the work in \cite{Lin:2022rbf,Lin:2023trc}. An important future direction is to extend these developments to the SUSY case, which was initiated by \cite{Boruch:2023bte}. Before commenting on the future direction, we provide remarks about the auxiliary Hilbert space construction. In this setting, one can build the physical Hilbert space with a one-particle insertion as
\begin{equation}
    \mH_{\rm phys}^{\rm 1p}:=\qty{\ket{n_L,n_R;AB,CD;j}}~,
\end{equation}
where $A,~B,~C,~D=\qty{X,O,\Omega}$, meaning that
\begin{equation}
    \ket{n_L,n_R;AB,CD;j}:=\ket{ABAB\cdots AB~\hmO_\Delta~CDCD\cdots CD;j}~,
\end{equation}
where the inner product in this Hilbert space has not been explicitly spelled out given that it involves a linear system of equations with several states. However, it should already be implicitly determined by the commutator relations in the chord algebra and the Hermitian conjugate operation \cite{Aguilar-Gutierrez:2025mxf}; although this should be confirmed explicitly.

We distinguish between two types of operators, depending on whether one adds matter chords to the left or the right
\begin{equation}
    \begin{aligned}
        \hmO_\Delta^{L}\ket{n,AB,j}:=\ket{0,n;\Omega,AB;j}~,\\
        \hmO_\Delta^{R}\ket{n,AB,j}:=\ket{n,0;AB,\Omega;j}~.
    \end{aligned}
\end{equation}
The two-sided Hamiltonian is defined in terms of the supercharges:
\begin{equation}
    \hH_L:=\qty{\hmQ_L,\hmQ_L^\dagger}~,\quad\hH_R:=\qty{\hmQ_R,\hmQ_R^\dagger}~,
\end{equation}
where
\begin{equation}
    \qty{\hmQ_i,\hmQ_j}=\qty{\hmQ^\dagger_i,\hmQ^\dagger_j}=0~,\quad \qty[\hH_L,\hH_R]=0~,
\end{equation}
for $i,j=\qty{L,R}$. By properly deriving the Hamiltonians with a matter insertion (which requires revisiting the Hamiltonians with one-particle inserted in \cite{Boruch:2023bte}), and finding the explicit inner products, we expect that one can evaluate Krylov operator or state complexity with particle insertions (similar to \cite{Ambrosini:2024sre,Aguilar-Gutierrez:2025pqp}), for instance using an initial state
\begin{equation}\label{eq:initial state}
    \ket{K_0}=\hmO_\Delta\ket{\Psi,j}~,
\end{equation}
where $\ket{\Psi,j}$ is the BPS HH state, and $\hmO_\Delta$ can be a BPS or non-BPS state, which we comment more about below. One could also study the evolution of \eqref{eq:initial state} to derive crossed four-point function (in terms of two-sided two-point functions, similar to \cite{Berkooz:2022fso,Aguilar-Gutierrez:2025pqp}). In the one-particle chord space, one might define general correlation functions (for BPS or half-BPS wormholes depending on $\hmO_\Delta$):\footnote{Taking the triple-scaling limit of these computations might allow the evaluate for the first time crossed-four point function for $\mathcal{N}=2$ super-Liouville theory with BPS states. I thank Jiuci Xu for pointing this out.}
\begin{equation}
    \bra{\Psi,j}\hmO_\Delta^\dagger\rme^{-\tau_L^*\hH_L-\tau_R^*\hH_R}q^{\Delta(\hn_X+\hn_O)_{\rm tot}}\rme^{-\tau_L\hH_L-\tau_R\hH_R}\hmO_\Delta^L\ket{\Psi,j}~.
\end{equation}
In particular, to work in the energy basis instead of chord basis to evaluate the chord inner product above, one has to generalize the \emph{entangler map} in bosonic DSSYK \cite{Okuyama:2024yya,Okuyama:2024gsn,Xu:2024gfm,Aguilar-Gutierrez:2025mxf} (which relates zero and one particle states) to the $\mathcal{N}=2$ case. The two-sided two-point functions can be then used to evaluate OTOCs \cite{Aguilar-Gutierrez:2025mxf}. It would be interesting to show whether the model is submaximally chaotic in the OTOC sense depending on the temperature, as found in the bosonic case \cite{Berkooz:2018jqr,Aguilar-Gutierrez:2025mxf,Aguilar-Gutierrez:2025pqp,Lin:2023trc}. 

\paragraph{BPS, Half-BPS \& non-BPS Wormholes}\label{ssec:1/2 BPS wormholes}
Assuming that $\hmO_\Delta$ is \textbf{not} a BPS operator, the BPS ``wormhole'' in \eqref{eq:BPS HH state j} with a particle insertion becomes:\footnote{I thank Adrián Sánchez-Garrido for suggesting adding matter to compare with the literature on fortuity.}
\begin{equation}
\begin{aligned}
\hmO^L_\Delta\ket{\Psi,j}&=\sum_{n=0}^\infty (\alpha_n\ket{0,n;\Omega,XO;j}+\beta_n\ket{0,n;\Omega,OX;j})~,
\end{aligned}
\end{equation}
which is a half-BPS state since
\begin{equation}
\begin{aligned}
        \hmQ_{R}^\dagger\hmO^L_\Delta\ket{\Psi,j}&=\hmQ_{R}\hmO^L_\Delta\ket{\Psi,j}=0~,\\
\hmQ_{L}^\dagger\hmO^L_\Delta\ket{\Psi,j}&\neq0~,\quad\hmQ_{L}\hmO^L_\Delta\ket{\Psi,j}\neq0~,
\end{aligned}
\end{equation}
given that $\hH_R\hmO^L_\Delta=\hmO^L_\Delta\hH_R$. This means that one can study the evolution of two-sided HH states of the form
\begin{equation}
    \rme^{-\tau_R\hH_R-\tau_L\hH_L}\hmO^L_\Delta\ket{\Psi,j}=    \rme^{-\tau_L\hH_L}\hmO^L_\Delta\ket{\Psi,j}~.
\end{equation}
Note that if we had inserted a BPS operator, i.e.~$\qty[\hat{V}_\Delta,\hmQ_{L/R}]=\qty[\hat{V}_\Delta,\hmQ^\dagger_{L/R}]=0$. This then leads to $\rme^{-\tau_R\hH_R-\tau_L\hH_L}\hat{V}_\Delta\ket{\Psi,j}= \hat{V}_\Delta\ket{\Psi,j}$, i.e.~the state is still BPS, and it would be again time independent. Meanwhile, if we inserted
\begin{equation}\label{eq:non_BPS state}
    \hmO_\Delta^L\hmO_\Delta^R\ket{\Psi,j}
\end{equation}
then \eqref{eq:non_BPS state} is no longer BPS,
\begin{equation}
    \hmQ_{L/R}\hmO_\Delta^L\hmO_\Delta^R\ket{\Psi,j}\neq0~,\quad\hmQ_{L/R}^\dagger\hmO_\Delta^L\hmO_\Delta^R\ket{\Psi,j}\neq0~.
\end{equation}
It would be interesting to do the evaluation of correlation functions and Krylov complexity for states and operators for the different combinations of BPS and non-BPS operators above. 

\subsection{Other outlook directions}\label{ssec:outlook}
\paragraph{Deforming Relational Holography} Gauge invariance in the boundary theory dual to any gauge-invariant gravity theory should be used as a fundamental principle to formulate holography in or outside the AdS/conformal field theory (CFT) correspondence. In particular, by deforming a boundary CFT to some other quantum field theory (dual to the bulk with a finite cutoff \cite{McGough:2016lol,Gross:2019ach,Gross:2019uxi,Kraus:2018xrn,Iliesiu:2020zld,Hartman:2018tkw,Zhang:2025dgm}, or with other boundary conditions \cite{Guica:2019nzm,Witten:2001ua,Parvizi:2025shq,Parvizi:2025wsg,Coleman:2020jte,Allameh:2025gsa,Ran:2025xas} or different background geometry \cite{Giveon:2017nie,Apolo:2019zai}), there is a corresponding gauge-invariant operator algebra whose observables change within the flow generated by a corresponding deformation parameter. In bulk terms, the observables are holographically described by different QRFs (due to modifications of the asymptotic boundary conditions) which affect the corresponding gauge-invariant observables. It would be interesting to investigate about relational holography in this scenario; particularly to connect finite cutoff thermodynamics with subsystem relational thermodynamics \cite{Hoehn:2023ehz}. We will approach this future direction with T$^2$ deformations in the bosonic DSSYK model in upcoming work.

\paragraph{The bulk theory dual of $\mathcal{N}=2$ DSSYK}
In this work, we developed a framework where the information about the boundary theory in a holographic system can be used to recover relational observables in its bulk dual, even though (i) there is no relational description of the boundary theory in the sense of the PW reduction map \cite{Page:1983uc}; and (ii) the precise bulk dual of $\mathcal{N}=2$ DSSYK beyond the low energy limit is currently unknown. One could carry out the reverse process from ours by starting from the bulk theory with a clock degree of freedom (such as an asymptotic boundary or some semiclassical feature where gravitational dressings can be defined). By interpreting the PW reduction map with the WDW constraint in terms of unitarily evolving physical states, one might associate a microscopic description of the system located on the same hypersurface as the clock internal degree of freedom. This might be a promising future direction to develop holography in more general spacetimes, where the boundary theory description is elusive, such as the static patch of dS space with a worldline observer.

Furthermore, one should explore the holographic correspondence in this system in more detail. For instance, we know from \cite{Penington:2024sum,Lin:2022zxd} the Lie superalgebra for the left/right boundary of $\mathcal{N}=2$ JT supergravity with matter. It would be interesting to derive the super-JT algebra from the super-chord algebra in \cite{Boruch:2023bte}. It would also be a next natural step to formalize the $\mathcal{N}=2$ super double-scaled algebra with respect to the bosonic case studied in \cite{Xu:2024hoc}. 

Besides the algebraic properties, one should check, similar to \cite{Fan:2021wsb} but for $\mathcal{N}=2$ super JT gravity, that the bulk length in the non-BPS HH state reproduces the corresponding bosonic answer, as our results from the boundary theory side indicate. Furthermore, one should be able to show that the triple-scaling limit (see App.~\ref{app:triple scaling limit}) of the $\mathcal{N}=2$ DSSYK Hamiltonian \cite{Boruch:2023bte} in an appropriate basis reproduces \eqref{eq:Hamiltonian N2 JT}.

Next, considering only the double-scaling limit with $\lambda\rightarrow0$, it is natural to ask whether one can find an explicit confirmation of our results from the holographic dual theory of the $\mathcal{N}=2$ DSSYK beyond the low energy regime (JT supergravity). One might expect that the same BPS wormhole lengths as \eqref{eq:BPS wormhole length} might be recovered in the HH preparation of state. Can (an extension of) the results in this work be used to derive the corresponding dual Hamiltonian? For instance, the Hamiltonian for $\mathcal{N}=2$ JT supergravity \cite{Lin:2022zxd} is
\begin{equation}\label{eq:Hamiltonian N2 JT}
    \hH=-\partial_\ell^2-\frac{1}{4}\partial_a^2+\rmi\qty([\hmQ_L^\dagger,\hel][\hmQ_R,\hel]\rme^{-\hel/2-\rmi \hat{a}}+[\hmQ_L,\hel][\hmQ_R^\dagger,\hel]\rme^{-\hel/2+\rmi \hat{a}})~,
\end{equation}
where $\hat{a}$ is a {gauge field} associated with the R-symmetry. It would be interesting to generalize this Hamiltonian for a q-deformed and UV finite generalization of JT supergravity, based on the bosonic case \cite{Blommaert:2024ymv}. We hope that this work can accelerate more progress in finding the bulk dual of $\mathcal{N}=2$ DSSYK based on recent proposals in the bosonic case that include complex Liouville string (sine dilaton gravity), and dS$_3$ \cite{Blommaert:2025eps} (see \cite{Anninos:2023exn} for an example of $\mathcal{N}=2$ dS$_2$ space).

\paragraph{More general systems}While, our proposal for BPS spread complexity (Sec.~\ref{ssec:BPS Krylov proposal}) was applied in a specific context (which we also implement for $\mathcal{N}=1$ DSSYK in App.~\ref{app:N1 wormholes}), we expect this can be used more generally in $\mathcal{N}=2$ quantum mechanics. The guiding principle to define BPS spread complexity \eqref{eq:BPS wormhole length} was the recurrence relation for the BPS coefficients in \eqref{eq:coefficients alpha beta} where each one can be used to define an effective Hamiltonian in a tridiagonal form. It would be interesting to deduce our proposal for more general systems with more diversity of BPS states while still obeying a recurrence relation with three terms that would allow one to formulate a tridiagonal effective Hamiltonian. A natural continuation of this work would be to apply our proposal in the $\mathcal{N}=2$ SYK at finite $N$ to make connection with the fortuity literature \cite{Chang:2024lxt}, which we discuss in the next paragraph. One should also try to carry out the lessons from this work to higher dimensions. For instance there has been interesting recent work on spread complexity in holographic supersymmetric models in \cite{Das:2024tnw}, where our notion of BPS spread might also provide new developments.

\paragraph{Fortuity}
Since we have an analytically solvable model, how does fortuity \cite{Chang:2024zqi,Chen:2024oqv,Chang:2024lxt} and related concepts (supercharge chaos, chaos invasion) manifest in it? These concepts, including the notions of monotone and fortuitous states and operators, have only been defined for finite $N$ systems, and fortuity is expected to be mostly manifested when $p\approx N/2$ (where $p$ is the number of all-to-all interactions, $N$ number of fermions, see the notation \eqref{eq:double scale}) in the large $N$ limit for the $\mathcal{N}=2$ SYK \cite{Chang:2024lxt}. Nevertheless, we expect the proposal for BPS spread complexity is related to the notions of BPS chaos \cite{Chen:2024oqv}, given that it reproduces a bulk observable holographically. It would be interesting to work in this context with R-charge concentration like (1.12) \cite{Chang:2024lxt} to show that BPS states with non-trivial boundary time flow are strongly (fortoitus) or weakly (monotonous) chaotic (that can be quantified by a large Thouless
time \cite{Chen:2024oqv}) by projecting operators onto appropriate subspaces even at $N\rightarrow\infty$. For instance, a comparison could be done by studying Krylov complexity for different  combinations of BPS and non-BPS operators in Sec.~\ref{ssec:particle space}. One might also try to develop a matrix model completion of the $\mathcal{N}=2$ DSSYK (such as a SUSY generalization of the eigenstate thermalization hypothesis (ETH) matrix model \cite{Jafferis:2022wez,Jafferis:2022uhu,Miyaji:2025ucp,Okuyama:2024eyf,Okuyama:2023aup}) to inquire more about the relationship with fortuity. Alternatively, one might propose a double-scaled generalization of fortuitous states \footnote{I thank Jiuci Xu for comments about this.} While the previous points are outside the scope of this work, we hope that this manuscript can spark new developments towards them.

\paragraph{York time for BPS states}Deducing the York Hamiltonian \cite{Parrikar:2025xmz} for $\mathcal{N}=2$ JT supergravity would be a useful calculation to investigate if BPS states may evolve in York time (which has been recently studied in bosonic JT gravity \cite{Parrikar:2025xmz}) in contrast to boundary time. One should begin deriving the Hamiltonian constraint from the ADM decomposition \cite{Arnowitt:1959eec,Arnowitt:1960es,Arnowitt:1960zza,Arnowitt:1960zzb,Arnowitt:1960zzc,Arnowitt:1961zz,Arnowitt:1961zza} using a constant mean curvature foliation in the supergravity action, and deducing the corresponding ADM Hamiltonian. Since there are additional gauge symmetries, they must be handled within the ADM framework, that may lead to an involved constraint analysis. We hope to report progress on this line of research in the future.

\paragraph{Entanglement entropy}
It was argued ever since \cite{Lin:2022zxd} that entanglement entropy in BPS wormholes can be negative if one considers a bulk theory with a large number of matter excitations, corresponding to a holographic dual theory at $N\rightarrow\infty$. It was recently proposed in \cite{Antonini:2025rmr} (see also \cite{Ouyang:2025cjr}) that the same observations hold even for non-SUSY black holes. The proposed resolution involves finite $N$ effects (higher genus contributions) to the  semi-quenched, quasi-quenched, and quenched Renyi entropy computations. In the DSSYK context, one similarly has that observables built from the chord algebra correspond to annealed ensemble-averaged observables of the physical SYK model in the double scaling limit \cite{Berkooz:2024lgq}. We expect that entanglement entropy of the BPS wormhole with BPS operator insertions will similarly lead to negative entropies due to annealed average, as well as for the non-BPS case. However, in this type of $N\rightarrow\infty$ there are additional terms in the relevant evaluations of entanglement entropy \cite{Goel:2023svz} which may lead to a positive entanglement entropy. Details about this are left for future work.

\section*{Acknowledgments}
I am grateful to Adrián Sánchez-Garrido and Jiuci Xu for initial collaboration. I thank Henry W. Lin for sharing a code to verify the super-chord algebra in \cite{Boruch:2023bte}; and specially to Gonçalo Araujo-Regado, Julian De Vuyst, Francesco Sartini and Jiuci Xu for providing useful feedback about a previous version of the draft; as well as Stefan Eccles, Philipp A. Höhn, Luca Marchetti and Josh Kirklin for useful discussions; and Andreas Blommaert for comments. I am thrilled to thank the organizers of the ``Quantum Reference Frames 2025'' conference in Okinawa Institute of Science where this work was presented and completed. SEAG is supported by the Okinawa Institute of Science and Technology Graduate University. This project/publication was also made possible through the support of the ID\#62312 grant from the John Templeton Foundation, as part of the ‘The Quantum Information Structure of Spacetime’ Project (QISS), as well as Grant ID\# 62423 from the John Templeton Foundation. The opinions expressed in this project/publication are those of the author(s) and do not necessarily reflect the views of the John Templeton Foundation.

\appendix
\section{Notation}\label{app:notation}
\paragraph{Acronyms}
\begin{itemize}[noitemsep]
\item ADM: Arnowitt-Deser-Misner formalism.
\item (A)dS: (Anti-)de Sitter.
\item BBNR: Berkooz-Brukner-Narovlansky-Rax.
\item BLY: Boruch-Lin-Yang.
\item BPS: Bogomol'nyi–Prasad–Sommerfield.
\item CFT: Conformal field theory.
\item CMC: Constant-mean-curvature.
\item (DS)SYK: (Double-scaled) SYK.
\item ETH: Eigenstate thermalization hypothesis.
\item HH: Hartle-Hawking.
\item JT: Jackiw-Teitelboim.
\item MSS: Momentum shift symmetry.
\item OTOC: Out-of-time-ordered correlator.
\item PW: Page-Wootters.
\item QRF: Quantum reference frame.
\item SUSY: Supersymmetry.
\item UV: Ultraviolet.
\item WDW: Wheeler-DeWitt.
\end{itemize}

\paragraph{Definitions}
\begin{itemize}
\item $N$, and $p$: Total number of fermions; and number of all-to-all interactions.
 \item $q=\rme^{-\lambda}:=\rme^{-\frac{p^2}{2N}}$: q-deformation parameter.
\item $(a;q)_n=\prod_{k=0}^{n-1}(1-aq^k)$: q-Pochhammer symbol.
\item $(a_1,a_2,\dots a_m;q)_n=\prod_{i=1}^N(a_i;~q)_n$.
\item $H_n(x|q)$ \eqref{eq:H_n def}: q-Hermite polynomials.
\item $|a,b):=\ket{a}\otimes\ket{b}$.
\item $E_{j_R}(\theta)$ \eqref{eq:energies N2}: Energy spectrum, where $\theta$ is a parametrization.
\item $\ket{v(\theta)}$, $\ket{u(\theta)}$ \eqref{eq:energy basis}: Energy basis. 
    \item $\mu(\theta)$ \eqref{eq:mu theta}: Energy basis measure.
\item $\hat{n}$, $\hat{P}$: chord number operator and its canonical conjugate.
\item $\hat{\ell}:=2\lambda \hat{n}$.
    \item $\hat{\mathcal{Q}}_{L/R}$, $\hat{\mathcal{Q}}_{L/R}^\dagger$ \eqref{eq:charges}: Supercharges.
    \item $\hH:=k\qty{\hat{\mathcal{Q}}_{L/R},\hat{\mathcal{Q}}_{L/R}^\dagger}$ \eqref{eq:n2 SYK hamiltonian}: $\mathcal{N}=2$ DSSYK Hamiltonian, with $k$ an arbitrary constant scaling.
    \item $\hJ_{L/R}$ \eqref{eq:dif operrators J}: R-charge generators.
    \item $j$: R-charge; and its rescaled form $j_R:=-j/2$.
    \item $\mathcal{H}_{\rm super-chord}$ \eqref{eq:physical state N2 jR}: Super-chord Hilbert space.
    \item $\ket{\Omega,j}$: Zero-chord state (maximally entangled state for fixed R-charge $j$).
        \item $\ket{n,XO,j}=\ket{XOXO\dots XO,j}$: Bosonic state from $n$ pairs of $XO$ operators.
        \item $\ket{n,OX,j}=\ket{OXOX\dots OX,j}$: Bosonic state from $n$ pairs of $OX$ operators.
    \item $\ket{n,XX,j}=\ket{XOXO\dots XOX,j}$: Fermionic state from $n$ pairs of $XO$ plus one $X$.
        \item $\ket{n,OO,j}=\ket{OXOX\dots OXO,j}$: Fermionic state from $n$ pairs of $OX$ plus $O$.
        \item $\ket{\Psi,j}$\eqref{eq:BPS HH state j}: BPS HH state.
        \item $\alpha_n$, $\beta_n$ \eqref{eq:alpha bbeta n}: BPS state coefficients.
        \item $\hH_{\rm WDW}$ \eqref{eq:WDW}: WDW constraint.
        \item $\hat{\Pi}_{\rm MSS}$ \eqref{eq:MSS projector}: MSS projector.
\item $\ket{t,j}$ \eqref{eq:ref clock state}: Clock states
\item $\chi(t-t')$ \eqref{eq:function chi}: Function of the boundary time difference in the clock inner product.
\item $\tilde{\ket{\Psi}}$ \eqref{eq:equiv class}: Equivalence class of kinematical states.
\item $\Pi_{\rm phys}$ \eqref{eq:curcial projector}: Coherent group averaging projector.
\item $ \mathcal{R}(\xi,j)$ \eqref{eq:reduction}: SUSY PW reduction map.
\item $\ket{\psi_{|R}(\xi,j)}$ \eqref{eq:perspective fixed}: Perspective-fixed Schrödinger state.
\item $\hat{\mathcal{O}}_R^{(\xi,j)}$ \eqref{eq:Gtwirl}: Relational observable dressed with respect to $R$ with a clock reading $t=\chi$ and R-charge $j$.
        \item $\hH_{\rm eff}$ \eqref{eq:alpha beta basis}: Effective Hamiltonian from the BPS coefficients.
        \item $\ell_*$ \eqref{eq:ini lenght}: Initial condition in the semiclassical wormhole distance.
    \item $\hat{\mathcal{L}}$ \eqref{eq:choi}: Choi–Jamiołkowski isomorphism.
    \item $\mathcal{C}_d$ \eqref{eq:BPS wormhole length}, $\hat{\mathcal{C}}_d$ \eqref{eq: def C_d}: BPS spread complexity, and the Krylov complexity operator.
    \item $\ket{K_n}$ \eqref{eq:Krylov basis}: Krylov basis.
    \item $a_n$, $b_n$ \eqref{eq:Krylov basis}: Lanczos coefficients.
\end{itemize}

\section{Complementary Background on \texorpdfstring{$\mathcal{N}=2$}{} DSSYK}\label{app:extra background}

In this appendix we complement the brief review about the $\mathcal{N}=2$ DSSYK in Sec.~\ref{sec:set up N2} starting from the finite $N$ model, its double-scaling limit, and constructing the auxiliary super-chord Hilbert space. However, we provide minimal additional details about this construction. For a detailed discussion this model, the reader is referred to the original works \cite{Boruch:2023bte,Berkooz:2020xne}.

\paragraph{Finite N System}Consider $N$ complex Majorana fermions, $\psi_i$, obeying
\begin{equation}
    \qty{\psi_i,\bar{\psi}_j}=\delta_{ij}~,\quad \qty{\psi_i,\psi_j}=0~.
\end{equation}
One can construct two supercharges in
\begin{equation}
    \hmQ_{\rm SYK}:=\sum_IC_I\Psi_I~,\quad \hmQ_{\rm SYK}^\dagger:=\sum_IC^*_I\overline{\Psi}_I~,
\end{equation}
where $\Psi_I:=\psi_{i_1}\cdots \psi_{i_p}$, and $C_I:=C_{i_1\dots i_p}$ are random couplings. The resulting Hamiltonian is built from the anticommutator of the supercharges:
\begin{equation}\label{eq:n2 SYK hamiltonian}
    \hH_{\rm SYK}:=k\qty{\hmQ_{\rm SYK},~ \hmQ_{\rm SYK}^\dagger}~,
\end{equation}
where $k$ is a constant with the same dimensions as energy, to keep the supercharges dimensionless. The R-charge generator in this model can be expressed as \cite{Berkooz:2020xne}
\begin{equation}
    \hat{J}_{\rm SYK}:=\frac{1}{2p}\sum_{i=1}^N\qty(\bar{\psi}_i{\psi}_i-{\psi}_i\bar{\psi}_i)~,
\end{equation}
so that $\hmQ$ has unit R-charge. In the following, we work with Gaussian distributed fermionic couplings (with normalization $\tr(H_{\rm SYK})=k$):
\begin{equation}
    \expval{C_I}_C=0~,\quad \expval{C_IC_{I'}^*}_C=\begin{pmatrix}
        N\\p
    \end{pmatrix}^{-1}2^p\delta_{I,I'}~,
\end{equation}
where the subindex $C$ indicates ensemble averaging over the couplings.

\paragraph{Double-Scaling Limit}Consider the double-scaling limit:
\begin{equation}\label{eq:double scale}
    N,~p\rightarrow\infty~,\quad\lambda:=\frac{2p^2}{N}~\text{ fixed},\quad q:=\rme^{-\lambda}\in[0,1)~.
\end{equation}
Following \cite{Berkooz:2020xne}, one can introduce a chord diagram where we label
\begin{equation}
    X:~\Psi_I~\text{nodes}~,\quad O:~\bar{\Psi}_I~\text{nodes}~.
\end{equation}
There are different Wick contractions between these operator strings that depend on the orientation between two given notes. In the terminology of \cite{Berkooz:2020xne}, we refer to a chord crossing where the contraction has the same/opposite orientation as a ``friend''/``enemy'' configuration. 

\paragraph{Auxiliary System}
We also replace the SYK supercharges and Hamiltonian by the ones of an auxiliary system
\begin{equation}\label{eq:op auxiliary}
   \hH_{\rm SYK}\rightarrow\hH~,\quad  \hmQ_{\rm SYK}\rightarrow\hmQ~,\quad \hat{J}_{\rm SYK}\rightarrow\hat{J}~,
\end{equation}
which can be used to build states within an auxiliary Hilbert space by first acting on a zero-chord state $\ket{\Omega}$ (similar to \cite{Berkooz:2018qkz,Berkooz:2018jqr}):
\begin{equation}
    \ket{O}:=\hmQ\ket{\Omega}~,\quad\ket{X}:=\hmQ^\dagger\ket{\Omega}~,
\end{equation}
and then appending the different combinations as (see \cite{Berkooz:2020xne} for more details)
\begin{equation}\label{eq:H aux}
    \mH_{\rm aux}=\bigoplus_{n=0}^\infty\qty{\ket{X},~\ket{O}}^{\otimes n}~.
\end{equation}
However, since $\hmQ$, $\hmQ^\dagger$ are  fermionic operators, one cannot have consecutive pairs $XX$ or $OO$ to construct non-trivial states. This means that the only states in the auxiliary Hilbert state have the form:
\begin{itemize}
    \item $\ket{n,XO}=\ket{XOXO\dots XO}$: Here, $n$ is the number of pairs of $XO$ operators (bosonic).
        \item $\ket{n,OX}=\ket{OXOX\dots OX}$: $n$ pairs of $OX$ operators (bosonic).
    \item $\ket{n,XX}=\ket{XOXO\dots XOX}$: $n$ pairs of $XO$ plus one $X$ (fermionic).
        \item $\ket{n,OO}=\ket{OXOX\dots OXO}$: $n$ pairs of $OX$ plus $O$ (fermionic).
\end{itemize}
We emphasize that the physical interpretation of the auxiliary states in the (super-)chord algebra is that they represent states within the physical bulk Hilbert space (which can be bosonic or fermionic) in contrast to states within the physical $\mathcal{N}=2$ SYK model in the double-scaling limit \cite{Lin:2022rbf,Berkooz:2020xne}.

Thus, $\hmQ^2=\qty(\hmQ^\dagger)^2=0$ implies that the surviving states (with respect to (\ref{eq:H aux})) have the form
\begin{equation}\label{eq:physical state N2}
    \qty{\ket{\Omega},~\ket{n,XO},~\ket{n,OX},~\ket{n,XX},~\ket{n,OO}}_{n=1}^\infty~.
\end{equation}
However, to generate BPS states, one needs an extension of the states with one-sided R-charge \cite{Boruch:2023bte}, which we turn to next.

\paragraph{Extending the super-algebra}
Here we introduce matter chords similar to the bosonic case, $\hmO_{\Delta}^{(L/R)}$, which is a double-scaled operator version of
\begin{equation}\label{eq:matter chord op}
    \hmO_\Delta:=\sum_IK_I\Psi_I~,\quad \Delta:=p'/p~,
\end{equation}
where $K_I:=K_{i_1\dots i_p}$ is another set of random couplings independent of $C_I$. This leads to a two-sided system where one can incorporate R-charge associated to each side in the construction of the Hilbert space, and it depends on the number of closed chords in the past (i.e.~forgotten friends and enemies in \cite{Boruch:2023bte}). We can then promote one-sided operators to two-sided ones (similar to \cite{Lin:2023trc}),
\begin{equation}
    \qty{\hmQ,~\hmQ^\dagger}\rightarrow\qty{\hmQ_L,~\hmQ_L^\dagger,~\hmQ_R,~\hmQ_R^\dagger}~,\quad \hJ\rightarrow\qty{\hJ_L,~\hJ_R}~.
\end{equation}
To simplify the evaluations within the zero-particle chord space \eqref{eq:physical state N2} (with an additional index denoting the R-charge), we consider $\Delta\rightarrow0$, so that we only need to work with one Hamiltonian $\hH$ instead of a two-sided system ($\hH_{L/R}$). This still leads to a $\mathcal{N}=4$ super-chord algebra (which can be extended with a two-sided Hamiltonian $\hH_{L/R}$ away from the $\Delta\rightarrow0$ limit \cite{Boruch:2023bte}),
\begin{subequations}
    \begin{align}
        \qty{\hmQ_i,~\hmQ_j}&=\qty{\hmQ_i^\dagger,~\hmQ_j^\dagger}=0~,\quad \qty{\hmQ_i,~\hmQ^\dagger_j}=\delta_{ij}\hH~,\label{eq:charges}\\
    [J_i,~\hmQ_j]&=\delta_{ij}\hmQ_j~,\qquad\qquad\quad [J_i,~\hmQ^\dagger_j]=-\delta_{ij}\hmQ^\dagger_j~,
    \end{align}
\end{subequations}
where $i=\qty{L,R}$. Since $[J_i,\hH]=0$, one has to include charges in labeling the states using the $U(1)_R$ generators
\begin{equation}
\begin{aligned}
    \qty(\hJ_R-\hJ_L)\ket{j}&=j\ket{j}~,\label{eq:dif operrators J}
\end{aligned}
\end{equation}
where  $j\in\mathbb{Z}$ is the R-charge, $\bra{j}\ket{j'}=\delta_{jj'}$, which allows to build states $\ket{n,AB,j}:=\ket{n,AB}\otimes\ket{j}$ 
with $A,B=\qty{X,~O}$, and similarly for $\ket{\Omega,j}$, which has the role of the maximally entangled state in $\mathcal{H}^j_{\rm phys}$ (\ref{eq:physical state N2}). This construction then leads to \eqref{eq:physical state N2 jR}.

\section{\texorpdfstring{Krylov Space of $\mathcal{N}=1$}{} Double-Scaled SYK}\label{app:N1 wormholes}
In this appendix we provide new results regarding spread complexity for BPS and non-BPS states which complement the discussion of the main text within the $\mathcal{N}=1$ DSSYK model. This model was introduced by \cite{Berkooz:2020xne},
\begin{equation}\label{eq:N1 theory}
    \hH_{\mathcal{N}=1}=k\hQ^2~,\quad\hQ=\hb+\hb^\dagger~,
\end{equation}
where $k$ is a constant, $\hb$ and $\hb^\dagger$ are fermionic creation and annihilation modes obeying an q-anticommutation relation $\qty{\hb,~\hb^\dagger}_q=1$, which act on a complete chord basis as\footnote{One can instead consider a multiplet representation of the wavefunctions of the same system that satisfy \cite{Belaey:2025ijg}
\begin{equation}
    -2\rmi\sin\theta\chi^\theta_{s}(n)=q^{1/2}\chi^\theta_{-s}(n-1)-(q^{-1/2}+s q^{n+1/2})\chi^\theta_{-s}(n+1)~,
\end{equation}
where $s=\pm$. The above recurrence relation uses a particular lightcone basis (thus the $\pm$ symbol); however, one can apply an appropriate change of basis to recover a similar form as the one reported in \cite{Berkooz:2020xne}\footnote{See (4.56) in \cite{Belaey:2025ijg}; and the discussion about the vanishing of unphysical modes in (4.57)}:
\begin{equation}
\cos\theta\xi_{\pm}^\theta(n)=\xi_{\mp}^\theta(n+1)+(1-(-q)^n)\xi_{\mp}^\theta(n-1)~,
\end{equation}
which corresponds to (\ref{eq:Q N1}).} \cite{Berkooz:2020xne}
\begin{equation}\label{eq:Q N1}
\hQ\ket{n}=k\qty(\ket{n+1}+(1-(-q)^n)\ket{n-1})~.
\end{equation}

\paragraph{Outline}In App.~\ref{sapp:BPS WH N1} we propose a natural extension of the Lanczos algorithm to characterize BPS states. In App.~\ref{sapp: Non BPS wormhole N1} we study the usual definition of Krylov complexity for non-BPS HH states, which we match to wormhole lengths in $\mathcal{N}=1$ JT supergravity in the semiclassical limit. This allows us to show a previous statement related to the complexity=volume conjecture in $\mathcal{N}=2$ JT supergravity \cite{Fan:2021wsb}.

\subsection{BPS States}\label{sapp:BPS WH N1}
Similar to Sec.~\ref{sec:BPS wormholes}, we construct a general BPS state in terms of the complete basis $\ket{n}$,
\begin{equation}\label{eq:BPS N1}
    \ket{\psi}=\sum_{n=0}^\infty a_n\ket{n}~,
\end{equation}
where $a_n$ are coefficients which satisfy the BPS constraint
\begin{equation}
    \hat{\mathcal{Q}}\ket{\psi}=\sum_n\qty(a_{n-1}+(1-(-q)^{n+1})a_{n+1})\ket{n}=0~.
\end{equation}
The recursion relation, 
\begin{equation}\label{eq: N1 recurrence relation}
    a_{n-1}+(1-(-q)^{n+1})a_{n+1}=0~.
\end{equation}
with initial condition $a_0=1$ is solved by the q-Hermite polynomials
\begin{equation}
    a_n=\frac{H_n(0|-q)}{(-q;-q)_n}~,
\end{equation}
where the first argument in the q-Hermite polynomial denotes the vanishing the energy eigenvalue in \eqref{eq: N1 recurrence relation}. Similar to the $\mathcal{N}=2$ case, we define spread complexity with respect to the initial reference state $a_0=1$, in terms of the Krylov basis $a_n$ as
\begin{equation}\label{eq:Krylov N1}
    \mathcal{C}_{\mathcal{N}=1}:=\sum_{n=0}^\infty n\abs{a_n}^2~.
\end{equation}
The above definition of Krylov complexity exactly reproduces the total chord number $\bra{\psi}\hat{n}\ket{\psi}$ in the BPS state \eqref{eq:BPS N1}. One can see that, in contrast to the $\mathcal{N}=2$ BPS case, there is no evolution in terms of R-charge in this system. This means that \eqref{eq:Krylov N1} is just a numerical constant, so it does not have a useful interpretation for the holographic dictionary. 

\subsection{Non-BPS}\label{sapp: Non BPS wormhole N1}
In this subsection, we consider the Hamiltonian \eqref{eq:N1 theory} in the basis \eqref{eq:Q N1},
\begin{equation}
\begin{aligned}
    \hH_{\mathcal{N}=1}\ket{n}=&k~\hmQ^2\ket{n}\\
    =&k\qty(\ket{n+2}+\qty(2+(-1)^n(q^{n-1}-q^n))\ket{n}+(1-q^n)(1-q^{n-1})\ket{n-2})~,
\end{aligned}
\end{equation}
and we study spread complexity \cite{Balasubramanian:2022tpr} for the non-BPS $\mathcal{N}=1$ HH state.

\paragraph{Krylov Basis}
We construct a natural Krylov basis starting from a reference $\ket{n=0}$, and define
\begin{equation}
\ket{K_n}:=\ket{2n}~,
\end{equation}
so that we can express the Hamiltonian as
\begin{equation}\label{eq:H N1 case}
\hH=k\qty(\rme^{-\rmi \hat{P}}+\qty(2+(q^{2\hn-1}-q^{2\hn}))+\rme^{\rmi \hat{P}}(1-q^{2\hn})(1-q^{2\hn-1}))~,
\end{equation}
where
\begin{equation}
    \rme^{\pm\rmi\hat{P}}\ket{K_n}=\ket{K_{n\pm1}}~,\quad\hn\ket{K_n}=n\ket{K_n}~.
\end{equation}
This can be used to study the following evolved state
\begin{equation}
    \ket{\psi(\tau)}=\rme^{-\tau\hH}\ket{K_0}=\sum_{n=0}^\infty\psi_n(\tau)\ket{K_n}~,
\end{equation}
where $\tau:=\rmi t+\frac{\beta}{2}$ and $\psi_n(\tau)=\bra{K_n}\rme^{-\tau\hH}\ket{K_0}$.

We can thus define the spread complexity of the HH state, corresponding to the choice of reference state in the above basis, by
\begin{equation}\label{eq:Krylov complexity N1}
    \mathcal{C}:=\eval{\frac{\bra{\psi(\tau)}\hat{n}\ket{\psi(\tau)}}{\bra{\psi(\tau)}\ket{\psi(\tau)}}}_{\tau=\frac{\beta}{2}+\rmi t}=\eval{\frac{\sum_nn\abs{\bra{\psi(\tau)}\ket{K_n}}^2}{\bra{\psi(\tau)}\ket{\psi(\tau)}}}_{\tau=\frac{\beta}{2}+\rmi t}~.
\end{equation}
We carry out the evaluation in the semiclassical limit below.\footnote{It would be interesting to relate our approach with recent work on path integral methods to approximate Krylov complexity \cite{Beetar:2025erl}.}
\paragraph{Semiclassical Evaluation}
Consider the path integral of the theory \eqref{eq:N1 theory},
\begin{equation}
   \int [\rmd\ell][\rmd P]\exp[\int\rmd\tau\qty(\frac{\rmi}{\lambda}P\partial_\tau\ell-H_{\mathcal{N}=1})]~.
\end{equation}
To find saddle point solutions, we work in the semiclassical limit $\lambda\rightarrow0$, with $\lambda n$ fixed, so that (\ref{eq:H N1 case}) reduces to
\begin{equation}
H_{\mathcal{N}=1}=k\qty(\rme^{-\rmi P}+2+\rme^{\rmi P}\qty(1-\rme^{-\ell})^2)~,
\end{equation}
where we label $\ell=2\lambda \expval{\hn}$.
Then, the saddle point are the solutions obey the following equations of motion
\begin{subequations}\label{eq:N1 EOM}
    \begin{align}
    \frac{1}{\lambda}\dv{\ell}{t}&=\pdv{H_{\mathcal{N}=1}}{P}=\rmi E(\theta)-{2k\rmi}\qty(\rme^{-\rmi P}+1)~,\\
    -\frac{1}{\lambda}\dv{P}{t}&=\pdv{H_{\mathcal{N}=1}}{\ell}={2k}\rme^{\rmi P-\ell}\qty(1-\rme^{-\ell})~.
\end{align}
\end{subequations}
To simplify the form of the above expressions, we let $k=J/\lambda$, so that \eqref{eq:N1 EOM} can be expressed as
        \begin{align}\label{eq:new EOM N1}
            \dv[2]{\ell}{t}=&4J^2~\rme^{-\ell}(1-\rme^{-\ell})~,\quad -\rmi\dv{\rme^{-\rmi P}}{t}=2J~\rme^{-\ell}\qty(1-\rme^{-\ell})~.
        \end{align}
The solution for $\ell(t)$ uses as the initial conditions,
\begin{subequations}\label{eq:N1 initial conditions}
    \begin{align}
    \bra{K_0}\rme^{-\frac{\beta}{2}\hH_{\mathcal{N}=1}}\hel\rme^{-\frac{\beta}{2}\hH_{\mathcal{N}=1}}\ket{K_0}&=\ell_*~,\\
    \eval{\dv{t}\bra{K_0}\rme^{-\tau^*\hH_{\mathcal{N}=1}}\hel\rme^{-\tau \hH_{\mathcal{N}=1}}\ket{K_0}}_{t=0}&=0~,
\end{align}
\end{subequations}
where $\ell_*$ is a constant determined by energy conservation, and $\tau:=\frac{\beta}{2}+\rmi t$. 

Given that the length $\ell$ has a positive acceleration \eqref{eq:new EOM N1} for $\ell_*>0$ in the initial conditions \eqref{eq:N1 initial conditions}, it follows $\rme^{-\ell}\approx0$ at late times. For these solutions, (\ref{eq:new EOM N1}) in the late time limit is approximated by
\begin{equation}
\begin{aligned}
    \ell(t)|_{t\gg J^{-1}}\approx{2J{\sin\theta}~t}+\ell_*~.
\end{aligned}
\end{equation}
Thus, spread complexity for the $\ket{n=0}$ state is given by
\begin{equation}\label{eq:spread N1}
    \mathcal{C}(t)|_{t\gg J^{-1}}=\eval{\frac{1}{2\lambda}\frac{\bra{\psi(\tau)}\hat{\ell}\ket{\psi(\tau)}}{\bra{\psi(\tau)}\ket{\psi(\tau)}}}_{t\gg J^{-1}}\approx\frac{J{\sin\theta}}{\lambda}t~.
\end{equation}
The above result is exactly as the bosonic DSSYK auxiliary system in this case \cite{Rabinovici:2023yex}. This is consistent with $\mathcal{N}=1$ super JT in \cite{Fan:2021wsb}. They found that  the semiclassical wormhole length (denoted $\mathcal{C}_{\rm V}$) at late times where quantum gravity is strongly coupled is still of the type
\begin{equation}
    \mathcal{C}_{\rm V}(t)\approx\frac{2\pi}{\beta_{\rm AdS}}t~,
\end{equation}
which indicates one should find the same bosonic result with fake temperature,
\begin{equation}
    \beta_{\rm AdS}=\frac{2\pi}{J\sin\theta}~.
\end{equation}
This confirms the conjectured complexity=volume \cite{Susskind:2014rva} in \cite{Fan:2021wsb}. The authors considered the wormhole length of the HH state in $\mathcal{N}=1$ JT gravity at leading order in the semiclassical approximation and for the disk topology. Our findings show that it matches the spread complexity of the non-BPS HH state \eqref{eq:Krylov complexity N1} in the late time regime.\footnote{We suspect the wormhole length and spread complexity match at all times, as in the bosonic DSSYK case \cite{Rabinovici:2023yex,Heller:2024ldz}; which one might confirm by revisiting the corresponding evaluation in \cite{Fan:2021wsb} from the bulk side, and \eqref{eq:new EOM N1} from the boundary one.}

\section{Dictionary Between the BLY and BBNR Basis}\label{app:BLY BBNR}
In the main text, we have mostly used the same notation for states and operators as Boruch-Lin-Yan (BLY) \cite{Boruch:2023bte}, while the normalization of states is  based on Berkooz-Brukner-Narovlansky-Raz (BBNR) \cite{Berkooz:2020xne} for convenience. In this short appendix we explain how to relate the normalizations from BLY \cite{Boruch:2023bte} (B.15-17) and those in this work (corresponding to BBNR \cite{Berkooz:2020xne} (4.38) with $s=-j/2$):
\begin{subequations}
    \begin{align}
        &\ket{\Omega,j}_{\rm BLY}=\ket{\Omega,j}_{\rm here}~,\\
        &\ket{n,OX,j}_{\rm BLY}=q^{\frac{n}{4}}\ket{n,OX,j}_{\rm here}~,\\
        &\ket{n,XO,j}_{\rm BLY}=q^{\frac{n}{4}}\ket{n,XO,j}_{\rm here}~,\\
        &\ket{n,OO,j}_{\rm BLY}=q^{\frac{n}{4}-\frac{j}{2}-\frac{1}{8}}\ket{n,OO,j}_{\rm here}~,\\
        &\ket{n,XX,j}_{\rm BLY}=q^{\frac{n}{4}+\frac{j}{2}-\frac{1}{8}}\ket{n,XX,j}_{\rm here}~.
    \end{align}
\end{subequations}
Meanwhile, the supercharges are related by
\begin{align}
    \hmQ_{\rm BLY}&=q^{-\frac{j}{2}+\frac{1}{8}}\hmQ_{\rm here}~,\quad \hmQ_{\rm BLY}^\dagger=q^{\frac{j}{2}+\frac{1}{8}}\hmQ_{\rm here}^\dagger~.
\end{align}

\section{\texorpdfstring{$\mathcal{N}=2$}{} DSSYK Partition function}\label{app:thermo}
In this appendix, we provide more details regarding the semiclassical thermodynamics of the $\mathcal{N}=2$ DSSYK (Sec.~\ref{sapp:semiclassical thermo}), which turn out to be very similar to those of the bosonic DSSYK model (see e.g.~\cite{Goel:2023svz}); and its triple-scaling limit (Sec.~\ref{app:triple scaling limit}).

Before the new results in the next subsections, we review known results in this introduction. As mentioned in the main text, the zero-chord state takes the role of the maximally entangled state for fixed R-charge $j$, which be used to define the HH state 
\begin{equation}
    \rme^{-\frac{\beta}{2}\hH}\ket{\Omega,j}~,
\end{equation}
and the partition function of the model at fixed $R$-charge,
or fixed chemical potential and temperature in the grand canonical ensemble respectively
\begin{equation}\label{eq:partition function}
    \begin{aligned}
        Z(\beta,\mu)&:=\frac{\bra{\Omega}\rme^{-\beta\hH-\mu \hat{J}_R}\ket{\Omega}}{\bra{\Omega}\ket{\Omega}}~,\quad Z(\beta,j):=\frac{\bra{\Omega,j}\rme^{-\beta\hH}\ket{\Omega,j}}{\bra{\Omega,j}\ket{\Omega,j}}~,
    \end{aligned}
\end{equation}
where $\ket{\Omega}:=\sum_{j}\ket{\Omega,j}$. To do explicit evaluations, we need to use the energy basis of the model to carry out the evaluations that depend on the basis in \eqref{eq:ortho bosonic basis}. We define
\begin{equation}\label{eq:energy basis N2}
    \begin{aligned}   
    &\ket{v(\theta)}=\sum_{n=0}^\infty\frac{q^{n/2}}{(q^2;q^2)_n}H_n(\cos\theta|q^2)\ket{H_n}~,\\
    &\ket{u(\theta)}=\sum_{n=0}^\infty\frac{q^{n/2}}{(q^2;q^2)_n}H_n(\cos\theta|q^2)\ket{\bar H_n}~,
    \end{aligned}
\end{equation}
where $H_n(x|q)$ is the q-Hermite polynomial
\begin{equation}\label{eq:H_n def}
    H_n(\cos\theta|q)=\sum_{k=0}^n\begin{bmatrix}
        n\\
        k
    \end{bmatrix}_q\rme^{\rmi(n-2k)\theta}~,\quad
    \begin{bmatrix}
        n\\
        k
    \end{bmatrix}_{q}:=\frac{(q;~q)_{n}}{(q;~q)_{n-k}(q;~q)_{k}}~.
\end{equation}
Using \eqref{eq:H N2 antisymmetric form} one gets
\begin{equation}\label{eq:energy basis}
    \begin{aligned}
     \hH\ket{v(\theta)}=E_{j_R}(\theta)\ket{v(\theta)}~,\quad\hH\ket{u(\theta)}=E_{-{j_R}}(\theta)\ket{u(\theta)}~,   
    \end{aligned}
\end{equation}
where the energy spectrum of this theory is
\begin{equation}\label{eq:energies N2}
    E_{j_R}(\theta)=k\Lambda_{j_R}(\theta):=2q^{-1/2}k\qty(\cosh(\lambda \qty(j_R-\frac{1}{2}))-\cos\theta)~,\quad \theta\in[0,\pi]~.
\end{equation}
The completeness relation of q-Hermite polynomials leads to \cite{Berkooz:2020xne}
\begin{equation}\label{eq:v compl}
    \bra{u(\theta')}\ket{u(\theta)}={q}^{-j_R}\Lambda_{-j_R}(\theta)\frac{2\pi}{(q^2,\rme^{\pm2\rmi\theta};q^2)_\infty}\delta(\theta-\theta')~,
\end{equation}
and we denoted $(a_1,a_2,\dots a_m;q)_n=\prod_{i=1}^N(a_i;~q)_n$. One also recovers a similar expression for $\bra{v(\theta)}\ket{v(\theta')}$ with $j_R\rightarrow-j_R$ in \eqref{eq:v compl}. From \eqref{eq:energy basis N2} and the relevant normalizations are\footnote{Note that although we are using the state notation in \cite{Boruch:2023bte}, the normalization is the one in \cite{Berkooz:2020xne} (4.37), instead of (B.15-17). One can also use the normalization in \cite{Boruch:2023bte}, there is a simple rescaling between them, App.~\ref{app:BLY BBNR}.} 
\begin{equation}\label{eq:inner product super chord}
    \begin{aligned}
         &\bra{n,XO,j}\ket{n',XO,j'}=\bra{n,OX,j}\ket{n',OX,j'}=q^{-n}(q^2;q^2)_{n-1}\delta_{nn'}\delta_{jj'}~,\\
    &\bra{n,OX,j}\ket{n',XO,j'}=-(q^2;q^2)_{n-1}\delta_{nn'}\delta_{jj'}~,\quad\bra{\Omega,j}\ket{\Omega,j'}=\delta_{jj'}.\end{aligned}
\end{equation}
One obtains the following wavefunctions
\begin{equation}\label{eq:inner product energy basis}
    \begin{aligned}
        &\bra{v(\theta)}\ket{H_n}=q^{j_R}\Lambda_{j_R}(\theta)H_n(\cos\theta|q^2)~,\\
        &\bra{u(\theta)}\ket{\bar H_n}=q^{-{j_R}}\Lambda_{-{j_R}}(\theta)H_n(\cos\theta|q^2)~,
    \end{aligned}
\end{equation}
which we apply in the following part.

\subsection{Semiclassical thermodynamics}\label{sapp:semiclassical thermo}
We now study the thermal properties of the system in the $\lambda\rightarrow0$ regime. For instance, using the previous relations, \eqref{eq:partition function} can be written as \cite{Berkooz:2020xne}
\begin{equation}\label{eq:full partition function of states}
    \begin{aligned}
   \bra{\Omega,j}\rme^{-\beta\hH}\ket{\Omega,j}=\int\rmd\theta~\mu(\theta)~\qty[q^{-j_R}\Lambda_{j_R}(\theta)^{-1}\rme^{-\beta E_{j_R}(\theta)}+q^{j_R}\Lambda_{-j_R}(\theta)^{-1}\rme^{-\beta E_{-j_R}(\theta)}]~.
    \end{aligned}
\end{equation}
where we denote the measure in energy basis as
\begin{equation}\label{eq:mu theta}
    \mu(\theta):=\frac{(q^2,\rme^{\pm2\rmi\theta};q^2)_\infty}{2\pi}~.
\end{equation}
For instance, for $q\rightarrow1^-$, this simplifies to
\begin{equation}
    \bra{\Omega,j}\rme^{-\beta\hH}\ket{\Omega,j}\eqlambda\int\frac{\rmd E(\theta)}{\Lambda_{j_R}(\theta)}~\rme^{S(\theta)-\beta E_{j_R}(\theta)}~,
\end{equation}
where in the semiclassical expression (which follows e.g.~\cite{Goel:2023svz}), $\Lambda_{-j_R}\eqlambda\Lambda_{j_R}$, and
\begin{align}
    S(\theta)&=S_0-\frac{(\frac{\pi}{2}-\theta)^2}{\lambda}~,
\end{align}
with $S_0$ an irrelevant overall constant, and while in the saddle point approximation,
\begin{equation}\label{eq:beta DSSYK}
    \beta\eqlambda\dv{S}{E_{j_R}}=-\frac{\pi-2\theta}{2J\sin\theta}~,
\end{equation}
where, again, we are using the parametrization \eqref{eq:parametrization k}. This means that $\theta=0$ corresponds to absolute zero temperature; while $\theta=\pi/2$ is the infinite temperature limit when $\lambda\rightarrow0$.

\subsection{Triple-Scaling Limit: Partition Function}\label{app:triple scaling limit}
In App D. \cite{Boruch:2023bte}, a triple-scaling for the $\mathcal{N}=2$ DSSYK was proposed so that one can recover the Schwarzian description of $\mathcal{N}=2$ JT supergravity by examining the BPS HH state $\ket{\Psi,j}$. The proposal is that
\begin{equation}\label{eq:TSL BLY}
\lambda\rightarrow0~,\quad \rme^{-2\lambda n}\rightarrow\frac{\rme^{-2\lambda n}}{(2\lambda)^2}~,
\end{equation}
where $q=\rme^{-\lambda}$ and $n$ is a label in the sums, such as for the $\mathcal{N}=2$ HH state \eqref{eq:BPS HH state j}. 

Next, we will explain how to recover the partition function of $\mathcal{N}=2$ JT supergravity \cite{Lin:2022zxd} from the the $\mathcal{N}=2$ DSSYK model \cite{Berkooz:2020xne}. From the bulk perspective:
\begin{equation}\label{eq:Z triple scaled SYK }
Z(\beta)=\rme^{S_0}\qty(\sum_{j_R=-1/2}^{1/2}\cos(\pi j)+\sum_{j}\int\rmd s~\rho(s)\frac{\rme^{-\beta E_{j_R}(s)}}{E_{j_R}(s)})~,
\end{equation}
where $S_0\in\mathbb{R}$ is a constant, and
\begin{equation}
    E_{j_R}(s)=s^2+\frac{1}{4}\qty(j_R-\frac{1}{2})^2~,\quad \rho(s)=\frac{2s\sinh(2\pi s)}{\pi}~.
\end{equation}
The first term in parenthesis in \eqref{eq:Z triple scaled SYK } comes from including $\sum_{j_R}\bra{\Psi,j}\ket{\Psi,j}$ for the HH state \eqref{eq:BPS HH state j} and evaluating the triple-scaling limit \eqref{eq:TSL BLY}.

From the boundary side, consider summing over the R-charge sectors in the DSSYK partition function $Z(\beta,j_R)$ (\ref{eq:full partition function of states})\footnote{The contribution from the BPS states in the partition function contained in the term $\cos(\pi j_R)$ in \eqref{eq:Z triple scaled SYK }; the details of the evaluation are in \cite{Boruch:2023bte}.}
\begin{equation}\label{eq:N2 partition function}
\begin{aligned}
    Z(\beta)&:=2\sum_{j_R}\int\rmd\theta~\frac{\left(q^2,e^{\pm2i\theta};q^{2}\right)_{\infty}}{2\pi}q^{-j_R}\Lambda_{j_R}(\theta)^{-1}\rme^{-\beta E_{j_R}(\theta)}\\
    E_{j_R}(\theta)&=k\Lambda_{j_R}(\theta)=2q^{-1/2}k\qty(\cosh(\lambda \qty(j_R-\frac{1}{2}))-\cos\theta)~,
\end{aligned}
\end{equation}
To carry out the triple-scaling limit in the partition function \eqref{eq:N2 partition function}, we propose to take
\begin{equation}
\theta=2\lambda ~s~,
\end{equation}
for real $s\sim \mathcal{O}(1)$ as $\lambda\rightarrow0$, and we rescale $\lambda \beta\rightarrow\beta$, considering that the overall proportionality constant in the energy as $k\sim\mathcal{O}(1/\lambda)$.
We then reproduce \eqref{eq:Z triple scaled SYK } with the know relation between the bosonic DSSYK energy measure in the triple-scaling limit (see e.g.~\cite{Berkooz:2024lgq}), namely
\begin{equation}
    \rmd\theta~\mu(\theta)\underset{\rm T.S.}{=}\frac{2\lambda(q^2;q^2)_\infty^3(1-q^2)^2}{2\pi}\rmd s ~\frac{2s}{\pi}\sinh(2\pi~s)~.
\end{equation}

\section{Alternative to BPS Spread complexity from \texorpdfstring{\eqref{eq:BPS HH state j}}{}}\label{app:alternative}
In this appendix, we study an alternative approach to define spread complexity associated to the Krylov basis recovered in Sec.~\ref{sec:BPS wormholes}.
The, arguably, most straightforward approach to defining complexity for the BPS HH state \eqref{eq:BPS HH state j} using the Krylov basis and the effective Hamiltonian in \eqref{eq:alpha beta basis} would be to consider the evolution of some reference state starting at $\ket{K_2}=\ket{B_0}$ with $b_2=0$,\footnote{We remind the reader there are no $\ket{K_{0}}$ and $\ket{K_{1}}$ states in this version of the Lanczos algorithm as explained in Sec.~\ref{sec:BPS wormholes}} and to evaluate the corresponding spread complexity. However, this would not probe the BPS state that we started with. This can be seen by constructing a state
\begin{align}\label{eq:recursion 1}
    \ket{\Psi_w}:=\rme^{-w\hH_{\rm eff}}\ket{B_0}=\sum_{n=2}^\infty\phi_n(w)\ket{K_n}~,\quad \phi_n(w):=\bra{K_n}\rme^{-w\hH_{\rm eff}}\ket{B_0}~,
\end{align}
where $w\in\mathbb{R}$ is some emergent time evolved by the effective Hamiltonian. One can try to evaluate the spread complexity of the above state as\footnote{Note that the coefficient in the spread complexity in \eqref{eq:exp val Krylov} is shifted $n\rightarrow n-2$ since in this case, the Lanczos algorithm begins at $n=2$; so that the definition translates to that in \cite{Balasubramanian:2022tpr}.}
\begin{equation}\label{eq:exp val Krylov}
    \frac{\sum_{n=2}^\infty (n-2)\abs{\bra{\Psi_w}\ket{K_n}}^2}{\bra{\Psi_w}\ket{\Psi_w}}~.
\end{equation}
A semiclassical approximation can be recovered by saddle point methods in the corresponding path integral (see also \cite{Xu:2024gfm,Aguilar-Gutierrez:2025hty,Aguilar-Gutierrez:2025pqp,Aguilar-Gutierrez:2025mxf})
\begin{equation}\label{eq:new path}
   \int[\rmd \ell][\rmd P]\exp\qty(\int\rmd w\qty(\frac{P}{\lambda}\partial_w\ell-H_{\rm eff}))~,\quad  H_{\rm eff}=\frac{J}{{\lambda}}\qty(\qty(1-\rme^{-\ell})\rme^{\rmi P}+\rme^{-\rmi P})~,
\end{equation}
where we take $J\in\mathbb{R}$ as an arbitrary constant, and $\lambda$ a small parameter in the semiclassical limit. We also defined
\begin{equation}\label{eq:length Bn}
    \hat{\ell}:=2\lambda \hn~,
\end{equation}
and we expressed expectation values in the state $\ket{\Psi_w}$
\begin{equation}
    \ell:=\bra{\Psi_w}\hat{\ell}\ket{\Psi_w}~,\quad P:=\bra{\Psi_w}\hat{P}\ket{\Psi_w}
\end{equation}
as fields in the path integral \eqref{eq:new path}.

The saddle point solutions obey the equations of motion
\begin{equation}\label{eq:EOM N2 effective}
    \frac{1}{\lambda}\dv{\ell}{w}=\pdv{H_{\rm eff}}{P}~,\quad -\frac{1}{\lambda}\dv{P}{w}=\pdv{H_{\rm eff}}{\ell}~.
\end{equation}
One can then deduce the initial conditions from the expectation values
\begin{subequations}\label{eq:initial cond N2 effective}
    \begin{align}
    &\bra{B_0}\hat{\ell}\ket{B_0}=2\lambda\bra{B_0}\hat{n}\ket{B_0}=0~,\\
    &\eval{\dv{w}\bra{B_0}\rme^{\rmi w\hH_{\rm eff}}\hat{\ell}\rme^{-\rmi w\hH_{\rm eff}}\ket{B_0}}_{w=0}=\rmi\bra{B_0}[\hH_{\rm eff},\hat{\ell}]\ket{B_0}=0~,
\end{align}
\end{subequations}
where both relations come from \eqref{eq:n hat Bn} and \eqref{eq:length Bn}. From \eqref{eq:EOM N2 effective} and the initial conditions in \eqref{eq:initial cond N2 effective} (i.e.~$\ell|_{w=0}=0$ and $\dv{w}\ell|_{w=0}=0$ in the semiclassical limit); one recovers
\begin{equation}\label{eq:ell semiclassical BPS}
    \ell(w)=2\log\cosh(J~w)~.
\end{equation}
Then, the semiclassical spread complexity, corresponding to the expectation value from \eqref{eq:length Bn} and \eqref{eq:ell semiclassical BPS} leads to the same answer as the spread complexity of the bosonic DSSYK \cite{Rabinovici:2023yex}. This is in sharp contrast to the expectation value of the semiclassical total chord number in BPS state/ the BPS wormhole length \eqref{eq:ell j}. However, the result is not surprising since the spread complexity associated with the $\ket{B_0}$ reference state does not need to be directly associated to \eqref{eq:BPS HH state j}.

\section{Details on Semiclassical Spread Complexity for Orthogonal Bosonic States}\label{app:detail ortho eval}
In this appendix, we show the details to recover \eqref{eq:complexity ortho}.
\paragraph{Expectation Values}
In the following, we study expectation values for the following states associated to $\mathcal{B} $ and $\bar{\mathcal{B}}$:
\begin{equation}\label{eq:psi tau}
   \ket{\psi(\tau)} =\begin{cases}
       \rme^{-\tau\hH}\ket{H_0}&\text{for }\mathcal{B}~,\\
       \rme^{-\tau\hH}\ket{\bar H_0}&~\text{for }\overline{\mathcal{B}}~,
   \end{cases}
\end{equation}
where $\tau:=\frac{\beta}{2}+\rmi t$ is a complexified time. One should note that \eqref{eq:psi tau} plays a natural role as the HH state in each of the subspaces \cite{Heller:2024ldz}. We also emphasize that the Hamiltonian \eqref{eq:H N2 operator} only acts on the states $\mathcal{B}\cup\bar{\mathcal{B}}$ and not $\ket{\Omega,j}$ (which we analyze in Sec.~\ref{ssec:TFD wormhole}), nor the ground (Sec.~\ref{sec:BPS wormholes}) and fermionic states. 

The initial conditions for the expectation value of the length operator are
\begin{subequations}\label{eq:initial conditions N2}
    \begin{align}\label{eq:ini lenght}
    \eval{\bra{\psi(\tau)}\hel\ket{\psi(\tau)}}_{\tau=\frac{\beta}{2}}&=\ell_*~,\\
    \eval{\dv{t}\bra{\psi(\tau)}\hel\ket{\psi(\tau)}}_{t=0}&=0~,\label{eq:iniitial 2}
\end{align}
\end{subequations}
where $\ell_*$ is a constant determined by energy conservation, and the second equality can be shown using the energy basis in (\ref{eq:energy basis N2}):
\begin{equation}\label{eq:integral towards proof}
    \begin{aligned}
        &\eval{\dv{t}\bra{\psi(\tau)}\hel\ket{\psi\qty(\tau)}}_{t=0}=\bra{\psi\qty(\frac{\beta}{2})}[\hH,\hel]\ket{\psi\qty(\frac{\beta}{2})}\\
        &\quad=\int\prod_{i=1}^2\rmd\theta_i\mu(\theta_i)
        \begin{cases}
            \bra{v(\theta_1)}\rme^{-\frac{\beta}{2}\hH}[\hH,\hel]\rme^{-\frac{\beta}{2}\hH}\ket{v(\theta_2)}&\text{for }\mathcal{B}~,\\
          \bra{u(\theta_1)}\rme^{-\frac{\beta}{2}\hH}[\hH,\hel]\rme^{-\frac{\beta}{2}\hH}\ket{u(\theta_2)}&\text{for }\overline{\mathcal{B}}~,
        \end{cases}
    \end{aligned}
\end{equation}
where we inserted the complete set of states $\ket{v(\theta)}$ or $\ket{u(\theta)}$ \eqref{eq:energy basis N2} for either $\qty{\ket{H_0},~\ket{\bar H_0}}$ as the initial state and we used (\ref{eq:inner product energy basis}). Note that:
\begin{equation}
    \begin{aligned}
        &\bra{v(\theta_1)}\rme^{-\frac{\beta}{2}\hH}[\hH,\hel]\rme^{-\frac{\beta}{2}\hH}\ket{v(\theta_2)}\\
    &=\rme^{-\frac{\beta}{2}\qty(E_{j_R}(\theta_1)+E_{j_R}(\theta_2))}\qty(E_{j_R}(\theta_1)\bra{v(\theta_1)}\hel\ket{v(\theta_2)}-E_{j_R}(\theta_2)\bra{v(\theta_1)}\hel\ket{v(\theta_2)})~,
    \end{aligned}
\end{equation}
so that we can perform a change of variables in (\ref{eq:integral towards proof}) $\theta_1\leftrightarrow\theta_2$ for the second term above, while keeping the first term the same, such that
\begin{equation}\label{eq:confirmation vanishing N2}
    \begin{aligned}
        &\bra{v(\theta_1)}\rme^{-\frac{\beta}{2}\hH}[\hH,\hel]\rme^{-\frac{\beta}{2}\hH}\ket{v(\theta_2)}\\
        &=E_{j_R}(\theta_1)\rme^{-\frac{\beta}{2}\qty(E_{j_R}(\theta_1)+E_{j_R}(\theta_2))}\qty(\bra{v(\theta_1)}\hel\ket{v(\theta_2)}-\bra{v(\theta_2)}\hel\ket{v(\theta_1)})\\
        &=\sum_{n=1}^\infty nE_{j_R}(\theta_1)\rme^{-\frac{\beta}{2}\qty(E_{j_R}(\theta_1)+E_{j_R}(\theta_2))} \qty(\bra{v(\theta_1)}\ket{H_n}\bra{H_n}\ket{v(\theta_2)}-\bra{v(\theta_2)}\ket{H_n}\bra{H_n}\ket{v(\theta_1)})~,
    \end{aligned}
\end{equation}
where we inserted the complete basis in (\ref{eq:Hn basis}) for $n\geq1$. However, we know that the inner products above are real from (\ref{eq:inner product energy basis}), which means that (\ref{eq:confirmation vanishing N2}) indeed vanishes, and \eqref{eq:iniitial 2} indeed follows.

\paragraph{Path integral Formulation}We study the $\mathcal{N}=2$ DSSYK path integral preparing the state \eqref{eq:psi tau} as
\begin{subequations}
\begin{align}
    &\int[\rmd P][\rmd \ell]\rme^{\int\rmd\tau\qty(\frac{\rmi}{\lambda} P\partial_\tau\ell-H)}~,\\
    &\text{where}~~H=q^{-1/2}k\qty(\rme^{-\rmi P}+\rme^{\rmi P}\qty(1-\rme^{-2\ell})+\qty(q^{-j_R+1/2}+q^{j_R-1/2}))~.\label{eq:conserved energy}
\end{align}
\end{subequations}
while $j_R\rightarrow-j_R$ for preparing $\rme^{-\tau\hH}\ket{\bar H_0}$. When $\lambda\rightarrow0$ and the rest is $\mathcal{O}(1)$, the saddle point must solve
\begin{align}
    &\frac{1}{\lambda}\dv{\ell}{t}=\pdv{H}{p}=2q^{-1/2}k\qty(\cos\theta-\rmi\rme^{-\rmi P})~,\\
    &\frac{1}{\lambda}\dv{P}{t}=-\pdv{H}{\ell}=-2q^{-1/2}k\rme^{-2\ell+\rmi P}~.
\end{align}
We can then combine the previous relations as:
\begin{equation}
    \frac{1}{\lambda^2}\dv[2]{\ell}{t}=4q^{-1}k^2\rme^{-\ell}~.
\end{equation}
Let us parametrize $E_{j_R}(\theta)$ in the same way as in (\ref{eq:energies N2}), and take the overall scaling as
\begin{equation}
    k=J~q^{1/2}/\lambda~.
\end{equation}
The initial conditions for the expectation values (\ref{eq:initial conditions N2}) in the classical fields above take the form
\begin{equation}
    \ell(t=0)=\ell_*~,\quad \eval{\dv{\ell}{t}}_{t=0}=0~.
\end{equation}
Then, the saddle-point solutions above are
\begin{subequations}\label{eq:solutions phase space N2}
    \begin{align}\label{eq:length path integral}
    \ell(t)&=\ell_*+2\log\cosh(J\sin\theta t)~,\\
    \rme^{-\rmi P(t)}&=\rmi\qty(\tanh(J\sin\theta t)-\cos\theta)~.
\end{align}
\end{subequations}
where $\ell_*$ is a constant determined by inserting (\ref{eq:solutions phase space N2}) in the conserved energy (\ref{eq:conserved energy}) with the parameterization (\ref{eq:energies N2}), namely
\begin{equation}\label{eq:ell times}
    \rme^{-\ell_*}=\sin^2\theta~.
\end{equation}
These results are applied in the main text to recover spread complexity in the $\lambda\rightarrow0$ limit, resulting in \eqref{eq:complexity ortho}.

\section{Details on Spread Complexity with Zero-Chord Reference State}\label{app:spread zero chord}
In this appendix, we provide additional details about the evaluation of \eqref{eq:complexity zero chord}.
\paragraph{Path Integral Evaluation}
To carry out the evaluation of the spread complexity using the zero-chord state as a reference state and the Hamiltonian representation \eqref{eq:Heritian Hamiltonian Omega N2}, we perform a canonical transformation
\begin{subequations}
\begin{align}
    \sqrt{(1-q^{\hn})(1+q^{\hn-1})}\rme^{-\rmi\hP}\rightarrow\rme^{-\rmi\hP}~,\\
    \rme^{\rmi \hP}\rightarrow\rme^{\rmi \hP}\sqrt{(1-q^{\hn})(1+q^{\hn-1})}~.
\end{align}
\end{subequations}
\eqref{eq:Heritian Hamiltonian Omega N2} then takes the form\footnote{The Hamiltonian takes a seemly non-hermitian form after performing this transformation. However, as explained in \cite{Aguilar-Gutierrez:2025mxf}, it remains Hermitian under the chord inner product \cite{Lin:2023trc,Xu:2024hoc} which is reflected on the commutation relations and the Hermitian conjugate operation of the chord algebra.}
\begin{equation}\label{eq:nonHeritian Hamiltonian Omega N2}
\begin{aligned}
    \hH'=\frac{J}{\lambda}\bigg(&\rme^{-\rmi \hP}+\rme^{\rmi \hP}(1-q^{\hn})(1+q^{\hn-1})+2\qty(\mathbb{1}\cosh\frac{\lambda}{2}+q^{\hn}\sinh\frac{\lambda}{2})\bigg)~.
\end{aligned}
\end{equation}
The path integral corresponding to (\ref{eq:nonHeritian Hamiltonian Omega N2}) becomes
\begin{subequations}\label{eq:path integral}
    \begin{align}
    &Z=\int[\rmd \ell][\rmd P]\rme^{\int\rmd\tau\qty(\frac{\rmi}{\lambda} P\partial_\tau\ell-H')}~,\\
     &H'=\frac{J}{\lambda}\qty(\rme^{-\rmi P}+\rme^{\rmi P}(1-\rme^{-2\ell})+2)~.\label{eq:Hprime N2}
\end{align}
\end{subequations}
The saddle point corresponds to the solution of
\begin{align}\label{eq:Other EOM Omega j}
    \frac{1}{\lambda}\dv{\ell}{t}&=\pdv{H'}{P}=\rmi E(\theta)-\frac{2J\rmi}{\lambda}\qty(\rme^{-\rmi P}+1)~,\\
    -\frac{1}{\lambda}\dv{P}{t}&=\pdv{H'}{\ell}=\frac{2J}{\lambda}\rme^{\rmi P-2\ell}~.
\end{align}
The scaling of the proportionality constant in the Hamiltonian is determined by fixing the normalization of traces in ensemble-averaged Hamiltonian moments of the with respect to the physical $\mathcal{N}=2$ SYK model in the double-scaling limit. From \eqref{eq:Other EOM Omega j} this leads us to
        \begin{align}
            \dv[2]{\ell}{t}=&2J^2~\rme^{-2\ell}~,\quad -\rmi\dv{\rme^{-\rmi P}}{t}=2J~\rme^{-2\ell}~.
        \end{align}
The solution for $\ell(t)$ then takes the form
\begin{equation}\label{eq:spread comple semiclassical}
\begin{aligned}
    &\ell(t)=2\log\frac{\cosh (J\sin\theta~t)}{\sin\theta}~,
\end{aligned}
\end{equation}
where we used as the initial conditions,
\begin{subequations}\label{eq:initial conditions N2 again}
    \begin{align}
    \bra{\Omega,j}\rme^{-\frac{\beta}{2}\hH'}\hel\rme^{-\frac{\beta}{2}\hH'}\ket{\Omega,j}&=\ell_*~,\\
    \eval{\dv{t}\bra{\Omega,j}\rme^{-\tau^*\hH'}\hel\rme^{-\tau\hH'}\ket{\Omega,j}}_{t=0}&=0~,\label{eq:new initial vel Omega j}
\end{align}
\end{subequations}
where $\ell_*$ is a constant determined by energy conservation and $\tau:=\frac{\beta}{2}+\rmi t$. Meanwhile, for \eqref{eq:new initial vel Omega j}, one has to evaluate 
\begin{equation}
    \eval{\dv{t}\bra{\Omega,j}\rme^{-\tau^*\hH'}\hel\rme^{-\tau\hH'}\ket{\Omega,j}}_{t=0}=\bra{\Omega,j}\rme^{-\frac{\beta}{2}\hH'}[\hH',\hel]\rme^{-\frac{\beta}{2}\hH'}\ket{\Omega,j}~,
\end{equation}
and for this one should insert a complete set of energy states. We can use the result that the Hamiltonian moments can be written as \cite{Berkooz:2020xne}:
\begin{subequations}
    \begin{align}\label{eq:energy integral}
  \bra{\Omega,j}\hH^n\ket{\Omega,j}=&k^n\int_0^\pi {\rmd\phi}~\mu(\phi) \left[q^{-{j_R}}(\Lambda_{j_R}(\phi))^{n-1} + q^{j_R}(\Lambda_{-{j_R}}(\phi))^{n-1} \right],\\
\mu(\phi):=&\frac{1}{2\pi}\left(q^2,e^{\pm2i\phi};q^{2}\right)_{\infty}~,
\end{align}
\end{subequations}
for $n\geq0$. As a consequence
\begin{equation}\label{eq:new eq initial condition Omega j}
\begin{aligned}
    &\bra{\Omega,j}\rme^{-\frac{\beta}{2}\hH'}[\hH',\hel]\rme^{-\frac{\beta}{2}\hH'}\ket{\Omega,j}\\
    &=\int\prod_{i=1}^2\rmd\theta_i\mu(\theta_i)\bigg(q^{-2j_R}\Lambda_{j_R}(\theta_i)^{-1}\bra{v(\theta_1)}\rme^{-\frac{\beta}{2}\hH}[\hH,\hel]\rme^{-\frac{\beta}{2}\hH}\ket{v(\theta_2)}\\
        &\hspace{3.5cm} +q^{2j_R}\Lambda_{-j_R}(\theta_i)^{-1}\bra{u(\theta_1)}\rme^{-\frac{\beta}{2}\hH}[\hH,\hel]\rme^{-\frac{\beta}{2}\hH}\ket{u(\theta_2)}\bigg)~.
\end{aligned}
\end{equation}
Using the same argument as (\ref{eq:confirmation vanishing N2}), we see that (\ref{eq:new eq initial condition Omega j}) vanishes, thus leading to (\ref{eq:new initial vel Omega j}). The initial conditions (\ref{eq:initial conditions N2 again}) and the conserved energy (\ref{eq:energies N2}) also lead again to $\rme^{-\ell_*}=\sin^2\theta$.

\section{Alternative Basis for \texorpdfstring{\eqref{eq:H N2 antisymmetric form_a}}{}}\label{app:alternative basis}
In this appendix, we complement the discussion in Sec.~\ref{ssec:TFD wormhole} by studying a basis orthogonal to \eqref{eq:Krylov basis N2 Omega} the Hamiltonian acting on the zero-chord state \eqref{eq:H N2 antisymmetric form_a} is tridiagonal as in \eqref{eq:Krylov basis}. This orthonormal basis was first noticed in \cite{Berkooz:2020xne},
\begin{equation}\label{eq:Ln basis}
    \ket{L_n}=\frac{\ket{n,OX,j}-\ket{n,XO,j}}{\sqrt{2q^{-n}(q^2;q^2)_{n-1}(1+q^n)}}~,
\end{equation}
where $\bra{L_n}\ket{L_m}=\delta_{nm}$. Using \eqref{eq:H N2 antisymmetric form}, one find that this does not generically lead to a tridiagonal Hamiltonian for all $j_R\in\mathbb{R}$ and $q\in[0,1)$,
\begin{equation}
\begin{aligned}
    \hH\ket{L_n}=&b_{n+1}\ket{L_{n+1}}+k\frac{q^{-j_R-1}+q^{j_R}-q^{n-j_R}+q^{n-j_R-1}}{\sqrt{2~q^{-n}(q^2;q^2)_{n-1}(1-q^n)}}\ket{n,OX,j}\\
    &+k\frac{q^{j_R+n}-q^{n+j_R-1}-q^{j_R-1}-q^{-j_R}}{\sqrt{2~q^{-n}(q^2;q^2)_{n-1}(1-q^n)}}\ket{n,XO,j}+b_n\ket{L_{n-1}}~.
\end{aligned}
\end{equation}
Nevertheless, there are special cases where we find a tridiagonal matrix,
\begin{equation}
    \hH\ket{L_n}=b_{n+1}\ket{L_{n+1}}+a_n\ket{L_n}+b_n\ket{L_{n-1}}~,
\end{equation}
corresponding to
\begin{itemize}
    \item $j_R=0$ and $q\in[0,1)$:
\begin{subequations}
    \begin{align}
    &a_n=k\qty(q^{-1}+1+q^{n-1}-q^{n})~,\\
     &b_{n}=k\sqrt{q^{-1}(1-q^{n-1})(1+q^n)}~.
    \end{align}
\end{subequations}
\item $j_R\sim\mathcal{O}(1)$ and $q\rightarrow1$:
\begin{subequations}
    \begin{align}
    &a_n=2k~,\\
     &b_{n}=k\sqrt{1-q^{2n}}~.
    \end{align}
\end{subequations}
\end{itemize}
However, since $\ket{L_n}=0$ is just empty, we chose to focus on $\ket{K_n}$ \eqref{eq:Krylov basis N2 Omega} in the main text.

\bibliographystyle{JHEP}
\bibliography{references.bib}
\end{document}